\documentclass[preprint]{JHEP3} 

\JHEPspecialurl{http://jhep.sissa.it/JOURNAL/JHEP3.tar.gz}
\usepackage{graphicx}
\usepackage{amsmath,amssymb}

\newcommand\fverb{\setbox\fverbbox=\hbox\bgroup\verb}
\newcommand\fverbdo{\egroup\medskip\noindent%
			\fbox{\unhbox\fverbbox}\ }
\newcommand\fverbit{\egroup\item[\fbox{\unhbox\fverbbox}]}
\newbox\fverbbox

\newcommand{\ep}{\epsilon}
\newcommand{\e}{\epsilon}
\newcommand{\nn}{\nonumber}
\newcommand{\ba}{\begin{equation}}
\newcommand{\ea}{\end{equation}}
\newcommand{\be}{\begin{eqnarray}}
\newcommand{\ee}{\end{eqnarray}}

\def\e{\epsilon}

\def\e{\epsilon}

\def\udcs{{\bar{u},d,\bar{c},s}}
\def\uscd{{\bar{u},s,\bar{c},d}}
\def\nn{\nonumber}

\title{\boldmath Next-to-leading order QCD predictions 
for $W^+W^+jj$ production  at the LHC
\unboldmath 
}
\author{Tom Melia \\Rudolf Peierls Centre for Theoretical Physics, 1 Keble Road, University of
   Oxford, UK\\
Email: \email{t.melia1@physics.ox.ac.uk}
}

\author{Kirill~Melnikov \\Department of Physics and Astronomy, Johns Hopkins 
University, Baltimore, MD 21218, USA\\
Email: \email{melnikov@pha.jhu.edu}
}

\author{Raoul R\"ontsch \\Rudolf Peierls Centre for Theoretical Physics, 1 Keble Road, University of
   Oxford, UK\\
Email: \email{r.rontsch1@physics.ox.ac.uk}
}
\author{Giulia Zanderighi \\Rudolf Peierls Centre for Theoretical Physics, 1 Keble Road, University of
   Oxford, UK\\
Email: \email{g.zanderighi1@physics.ox.ac.uk}
}

\received{\today} 		
\accepted{\today}		

\abstract{Because the LHC is a proton-proton collider, sizable
  production of two positively charged $W$-bosons in association with
  two jets is possible. This process leads to a distinct signature of
  same sign high-$p_{\perp}$ leptons, missing energy and jets. We
  compute the NLO QCD corrections to the QCD-mediated part of $pp \to
  W^+W^+jj$. These corrections reduce the dependence of the production
  cross-section on the renormalization and factorization scale to
  about $\pm 10$ percent. We find that a large number of $W^+ W^+ jj$
  events contain a relatively hard third jet.  The presence of this
  jet should help to either pick up the $W^+ W^+ jj$ signal or to
  reject it as an unwanted background.}


\begin{document} 

\section{Introduction} 
The large energy and large luminosity of the LHC lead to sizable
production cross-sections for multi-particle final states, including
rather exotic ones. For example, one can produce a pair of positively
charged $W$-bosons, in association with two jets $pp \to W^+ W^+
jj$. At $\sqrt{s} = 14~{\rm TeV}$, the cross-section for this process
is about $1~{\rm pb}$ and therefore accessible. Leptonic decays of
$W$-bosons give rise to two positively charged isolated leptons and
missing energy, which is nearly a background-free signature.

The observation of this process is interesting in its own right, but
there are other reasons to study it. Of particular importance are
various physics cases for which $pp \to W^+ W^+ jj$ is a background
process.  Interestingly, such cases can be found both within and
beyond the Standard Model.  For example, it is possible to use
same-sign lepton pairs to study {\it double parton scattering} at the
LHC \cite{dps} in which case the single scattering process $pp \to W^+
W^+ jj$ is the background. Events with same-sign leptons, missing
energy and two jets can also appear due to resonant slepton production
which may occur in $R$-parity violating SUSY models \cite{dreiner} or
in the case of diquark production \cite{han} with subsequent decay of
the diquark to e.g. pairs of top quarks. Similarly, one of the
possible production mechanisms of the double-charged Higgs boson at
the LHC has a signature of two same-sign leptons, missing energy and
two jets \cite{maalampi}.

At leading order in the perturbative expansion in QCD, the $W^+ W^+
jj$ final state is produced in proton collisions by both electroweak
(EW) and QCD mechanisms.
Interference terms between these mechanisms are 
doubly suppressed due to the different color structure of the two 
processes; interference only occurs at sub-leading
color and even then only when the quarks are all identical. We 
neglect these terms and consider the QCD process separately.
The corresponding cross-sections scale as $\sigma_{\rm EW}
\sim \alpha_{\rm EW}^4$ and $\sigma_{\rm QCD} \sim \alpha_{\rm EW}^2
\alpha_s^2$, where $\alpha_{\rm EW}$ and $\alpha_s$ are electroweak
and strong coupling constant respectively. Given the large hierarchy
between strong and weak coupling constants, one expects $\sigma_{\rm
  QCD} \gg \sigma_{\rm EW}$, but this turns out to be too naive. In
reality, the production cross-section due to a gluon exchange is only
about fifty percent larger than the production cross-section of $pp
\to W^+ W^+ jj$ by electroweak mechanisms \cite{dps}.

As we mentioned earlier, $W^+ W^+ jj$ production is the background to
a number of interesting beyond the Standard Model (BSM) physics
processes. It is peculiar that NLO QCD corrections to some of these
{\it signal} BSM processes were calculated, see
e.g. Refs.~{\cite{dreiner,han}}.  Also, the NLO QCD corrections to the
{\it electroweak} production of $W^+ W^+ jj$ were calculated recently
in Ref. {\cite{jager}}, but similar corrections to the QCD production
of $W^+ W^+ jj$ are unknown. There is a clear reason for this: the
computation of NLO QCD corrections to the QCD-induced process $pp \to
W^+ W^+ jj$ is more involved because $pp \to W^+ W^+ jj$ is a $2 \to
4$ process.  As a result, the NLO QCD calculation for QCD-induced $pp
\to W^+ W^+ jj$ requires dealing with one-loop six-point tensor
integrals of relatively high rank, while for the EW-induced process
this is not the case since at Born level there is no color exchange
between the quark lines. Until very recently, theoretical methods for
one-loop calculations were inadequate to handle computations of such a
complexity, but the situation has changed dramatically in the past two
years.  Thanks to recent technical developments
\cite{denner,golem,britto,Britto:2004tx,Forde:2007mi,opp,opp2,egk,Giele:2008ve,cfb},
NLO QCD computations for such $2 \to 4$ processes as $pp \to W (Z) +
3~{\rm jets}$, $pp \to t \bar{t} + b \bar{b}, q \bar q \to b \bar b b
\bar b$ and $pp \to t \bar{t} + 2~{\rm jets}$, have been performed
during the {\it past} year \cite{Bredenstein:2009aj,
  Bredenstein:2010rs, Bevilacqua:2009zn,
  Berger:2009zg,Berger:2009ep,Ellis:2009zw,KeithEllis:2009bu,
  Melnikov:2009wh,bbbb,tt2j,Berger:2010vm}.  Similar techniques should
be applicable to NLO QCD computations of processes with several
electroweak gauge bosons and jets, and $pp \to W^+W^+jj$ is an
interesting example.  We also note that the computation of NLO QCD
corrections to $pp \to W^+W^+jj$ involves a small subset of amplitudes
needed for the computation of NLO QCD corrections to $pp \to W^+ W^-
jj$, which is an important background to Higgs boson production in
weak boson fusion.

The remainder of the paper is organized as follows.  In
Section~\ref{sc2} we briefly discuss technical aspects of the
calculation.  In Section~\ref{sc3} we describe the results. In
Section~\ref{sc4} we present our conclusions.  In the Appendix,
numerical results for one-loop helicity amplitudes for $0 \to (
\bar{q}_i W^+ q_j) ( \bar{q}_k W^+ q_m)$ are given.

\section{Technical details} 
\label{sc2}

In this Section, technical aspects of the computation are
summarized. It is well-known that the computation of NLO QCD
corrections to any process requires three ingredients -- one-loop
virtual corrections to a Born process, real emission corrections and
the subtraction counter-terms for infra-red and collinear
singularities.  To compute the one-loop virtual corrections, we use
the framework of generalized $D$-dimensional unitarity
{\cite{egk,Giele:2008ve}}, closely following and extending the
implementation described in Ref.~{\cite{Ellis:2008qc}}.  Because the
one-loop virtual corrections are computed using unitarity cuts, an
important ingredient for the computation of one-loop corrections are
the tree-level helicity amplitudes $0 \to ( \bar{q}_i W^+ q_j) (
\bar{q}_k W^+ q_m) + g$. We compute those helicity amplitudes using
Berends-Giele recursion relations \cite{BG}.  Incidentally, these
amplitudes are the ones that are needed to compute the real emission
corrections to $pp \to W^+ W^+ jj$.  The subtraction terms are
calculated using the Catani-Seymour dipole formalism \cite{cs}. We
employ the optimization of the subtraction technique suggested in
Ref.~{\cite{nagy}}, where the subtraction is only performed if the
kinematics of the event is close to either soft or collinear
limit. Our implementation of the Catani-Seymour formalism closely
follows the program MCFM {\cite{mcfm}}.  In fact, we use the MCFM
framework extensively to combine virtual and real emission corrections
and the subtraction terms. Finally, we evaluate one-loop scalar
integrals using the {\sf QCDloop} library \cite{qcdloops}.

We now describe the computation in some detail.  First, we point out
that within the generalized unitarity framework, a basic object that
needs to be calculated for each phase-space point is a one-loop
helicity amplitude. To decouple color, helicity amplitudes are
expressed through primitive amplitudes~\cite{Bern:1994fz}.
Those primitive amplitudes can be computed with the help of the
color-stripped Feynman rules~\cite{Bern:1994fz,Bern:1997sc,rules}.  We
note that, for the calculation of a given primitive amplitude,
color-charged particles are ordered while all permutations of
color-neutral particles must be considered, to achieve a
gauge-invariant result.  For our purposes, this implies that the
ordering of the $W^+$ bosons is not fixed and we have to account for
all possible insertions of the $W^+$ bosons. Fortunately, given the
fact that both of the two $W$ bosons have the same charge, this is a
relatively minor complication since two $W$-bosons can not couple to
the same quark line. Throughout the paper we treat the top quark as
infinitely heavy, while all other quarks are taken to be massless.  
We do not consider final state top production as this leads to a different
experimental signature. Top quarks therefore only contribute through fermion
loops. While it is possible to deal with massive particles in the loop with D-dimensional
unitarity~\cite{Ellis:2008ir,Melnikov:2010iu}, we choose to neglect these effects since
the momentum transfer through the fermion loop is far
below the threshold for top pair production. This is consistent with the treatment of parton distribution functions,
which use $\beta_0$ with five flavours in the evolution of $\alpha_s$.

\begin{figure}[t]
\begin{center}
\includegraphics[angle=0,scale=1]{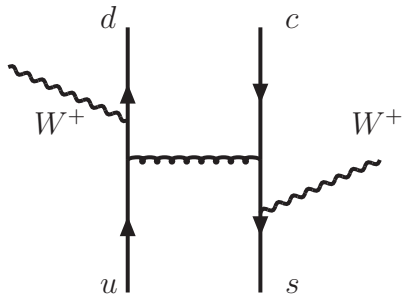}
\caption{A typical Feynman diagram that contributes to QCD production
  of $W^+W^+jj$ in hadron collisions at leading order in perturbative
  QCD.}
\label{fig0}
\end{center}
\end{figure}

We now give the decomposition of the scattering amplitudes in terms of
primitive amplitudes.  We begin with the tree-level process $0 \to
\bar u\,d\,\bar c\,s \, + (W^+ \to e^+\nu_e)+(W^+\to \mu^+\nu_\mu)$,
where we fix the flavor structure, for definiteness and treat the
Cabibbo-Kobayashi-Maskawa matrix as diagonal.
The amplitude for this process reads
\begin{eqnarray}
&&A^{\rm tree}(\bar u, d, \bar c, s; e^+, \nu_e, \mu^+, \nu_\mu) =
g_s^2 \left(\frac{g_W}{\sqrt{2}}\right)^4 
P_{W^+}(s_{e^+ \nu_e})
P_{W^+}(s_{\mu^+ \nu_\mu})
\nonumber \\
&&
\times \left ( \delta_{i_ d i_{\bar c}} \delta_{i_s i_{\bar u}} 
- \frac{1}{N_c} \delta_{i_d i_{\bar u}} \delta_{i_s i_{\bar c}} 
\right ) A_0(\bar u, d, \bar c, s), 
\label{eq0}
\end{eqnarray}
where $g_{s,W}$ are the strong and weak coupling constants,
respectively, quark color indices are indicated explicitly, $N_c = 3$
is the number of colors, and lepton labels in the right hand side of
Eq.(\ref{eq0}) are suppressed. Also, we use the $W^+$ propagators with
a Breit-Wigner form in Eq.(\ref{eq0})
\begin{equation}
P_{W^+}(s)\equiv \frac{s}{s-M_W^2 + i \Gamma_W M_W}\,.   
\end{equation}
Finally, to account for leptonic decays $W^+ \to l^+(q_1) \nu_l(q_2)$
in Eq.(\ref{eq0}), we replace the polarization vector of the outgoing
$W^+$ bosons by \ba \epsilon_{-}^\mu (q_1,q_2) = \frac{\bar{u}(q_2)
  \gamma_\mu \gamma_{-} v(q_1)} {(q_1+q_2)^2},\;\; \gamma_{-} =
\frac{1 - \gamma_5}{2}.
\label{eq:wpol}
\ea
A typical diagram that contributes to the amplitude $A_0(\bar u,d,\bar
c, s)$ is shown in Fig.~\ref{fig0}.

The computation of real emission corrections requires the scattering
amplitude for $0 \to \bar u\, d\, \bar c\, s\, + g+ (W^+ \to
e^+\nu_e)+(W^+\to \mu^+\nu_\mu)$.  The amplitude for this process is
written in terms of primitive amplitudes
\begin{eqnarray}
&& A^{\rm tree}(\bar u,  d, \bar c, s, g; e^+, \nu_e,\mu^+, \nu_\mu) = 
g_s^3 \left(\frac{g_W}{\sqrt{2}}\right)^4 
P_{W^+}(s_{e^+ \nu_e})
P_{W^+}(s_{\mu^+ \nu_\mu})
\nonumber  \\ && 
\qquad \times \Bigg [ 
T^{a}_{i_s i_{\bar u}} \delta_{i_d i_{\bar c}} A_0(\bar u, d, \bar c,  s, g) 
+ T^{a}_{i_d i_{\bar c}} \delta_{i_s i_{\bar u}} A_0(\bar u, d, g, \bar c, s) 
\nn \\ && 
\qquad+\frac{1}{N_c}\left(
T^{a}_{i_d i_{\bar u}} 
\delta_{i_s i_{\bar c}} A_0(\bar u, g, d, \bar c, s) 
+ T^{a}_{i_s i_{\bar c}} \delta_{i_d i_{\bar u}} A_0(\bar u, d, 
\bar c,g, s)\right)
\Bigg ]
 .
\end{eqnarray}
Again, in the right hand side of this equation the lepton labels are
suppressed.  Finally, the one-loop amplitudes required for this
calculation can be written as
\begin{eqnarray}
&&A^{\rm one-loop}(\bar u,d, \bar c, s;e^+, \nu_e, \mu^+, \nu_\mu) =
\nonumber \\
&&\qquad g_s^4 \left(\frac{g_W}{\sqrt{2}}\right)^4 
P_{W^+}(s_{e^+ \nu_e})
P_{W^+}(s_{\mu^+ \nu_\mu})
\left [ \delta_{i_ d i_{\bar c}} \delta_{i_s i_{\bar u}} A_1
+ \delta_{i_d i_{\bar u}} \delta_{i_s i_{\bar c}}  A_2 \right ]\,.
\label{eq1}
\end{eqnarray}

\begin{figure}[t]
\begin{center}
\hspace*{1cm}
\includegraphics[angle=0,scale=1]{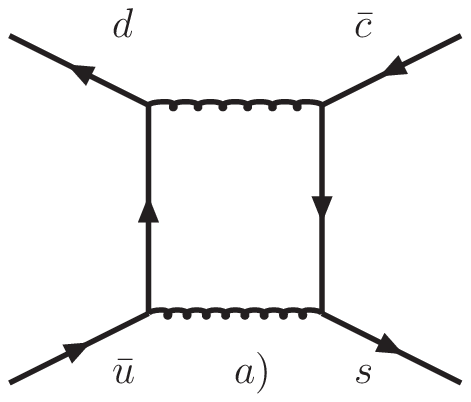}\hspace*{1cm} 
\includegraphics[angle=0,scale=1]{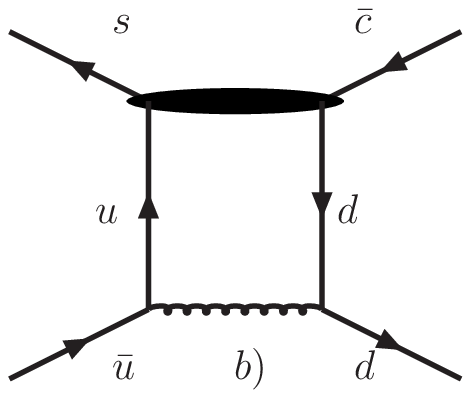}\\
\vspace*{1cm}
\hspace*{1.0cm}\includegraphics[angle=0,scale=1]{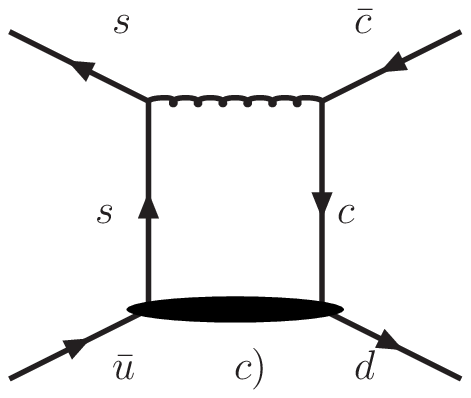}\hspace*{1cm}
\includegraphics[angle=0,scale=1]{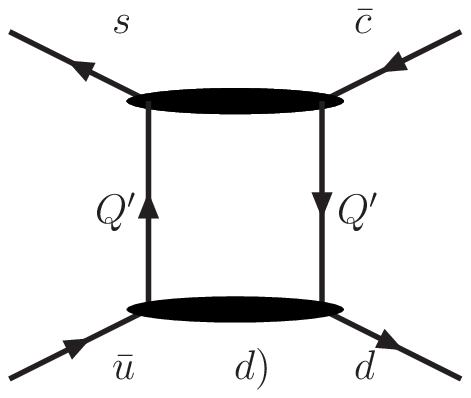}
\caption{Parent diagrams for one-loop primitive amplitudes 
for $0 \to \bar u d \bar c s+W^+W^+$. Shaded areas represent dummy lines 
which can not be cut.  $W^+$ bosons are not shown. Quarks $Q'$  
do not couple to $W$-bosons. }
\label{fig0a}
\end{center}
\end{figure}

The color-ordered amplitudes $A_{1,2}$ can be expressed through the
primitive amplitudes shown in Fig.~\ref{fig0a}.  The relations read
\begin{eqnarray}
 A_1 &=& 
\left(N_c-\frac{2}{N_c}\right) A_a(\bar u, d, \bar c, s)
-\frac{2}{N_c} A_a(\bar u, d, s, \bar c)
\nonumber\\ 
& 
-&\frac{1}{N_c} A_b(\bar u, s, \bar c, d)
-\frac{1}{N_c} A_c(\bar u, s, \bar c, d)
+ n_f A_d(\bar u, d, \bar c, s) 
\,,
\label{a1}\\ 
 A_2 &=& \frac{1}{N_c^2} A_a(\bar u, d, \bar c,s)
+\left(1+\frac{1}{N_c^2}\right) A_a(\bar u, d, s,\bar c)
\nonumber
\\
& +&\frac{1}{N_c^2} A_b(\bar u, s, \bar c, d)
+\frac{1}{N_c^2} A_c(\bar u, s, \bar c, d)
-\frac{n_f}{N_c} A_d(\bar u, d, \bar c, s) \,.
\label{a2}
\end{eqnarray}
Amplitudes with two adjacent quarks can be reduced to amplitudes where
quarks and anti-quarks alternate using the C-parity relation
\begin{equation}
A_a(\bar u^{\lambda_{u}}, d^{\lambda_{d}}, s^{\lambda_{s}}, \bar c^{\lambda_{c}}) = - 
A_a(\bar u^{\lambda_{u}}, d^{\lambda_{d}}, \bar s^{\lambda_{s}}, c^{\lambda_{c}})\,, 
\label{eq2.8}
\end{equation}
where we explicitly indicate fermion helicities\footnote{We note that
  the r.h.s. of Eq.~(\ref{eq2.8}) is non-zero, since once the
  helicities of the fermions are fixed, we implement the coupling of
  the $W$-bosons to fermions as a vector coupling.}.

Amplitudes with two identical quarks and anti-quarks are obtained from
the amplitudes shown above by anti-symmetrising quarks or anti-quarks
and including the appropriate symmetry factors in the cross-section
calculation.  Finally, we note that numerical results for helicity
amplitudes at a particular phase-space point are given in the
Appendix.

\section{Results}
\label{sc3}

In this Section, we present the results of the calculation.  We
consider proton-proton collisions with the center-of-mass energy
$\sqrt{s} = 14~ {\rm TeV}$. We require leptonic decays of the
$W$-bosons and consider $e^+ \mu^+ \nu_e \nu_\mu $ final state. The
$W$-bosons are on the mass-shell and we neglect quark flavour mixing.
We note that if identical leptons in the final state are present,
there are interference effects not included in our
calculation. However, such interference effects force the $W$-bosons
off the mass shell, so that their numerical importance is
limited. Within this approximation, the cross-sections for same- and
different-flavor production are related $\sigma (pp \to \mu^+ \mu^+ X)
= \sigma (pp \to e^+ e^+ X) = 0.5\, \sigma (pp \to e^+ \mu^+X)$. This
implies that the cross-section for the full flavor sum $e^+ \mu^+ +
e^+ e^+ + \mu^+ \mu^+$ can be obtained by multiplying our results by a
factor two.

We impose standard cuts on lepton transverse momenta $p_{\perp, l} >
20~{\rm GeV}$, missing transverse momentum $p_{\perp, \rm miss} >
30~{\rm GeV}$ and charged lepton rapidity $| \eta_l| < 2.4$. We define
jets using anti-$k_{\perp}$ algorithm \cite{Cacciari:2008gp}, with
$\Delta R_{j_1j_2} = \sqrt{(\eta_{j_1} - \eta_{j_2})^2 + (\phi_{j_1} -
  \phi_{j_2})^2} =0.4$ and, unless otherwise specified, with a
transverse momentum cut $p_{\perp, j} = 30~{\rm GeV}$ on the two
jets. The mass of the $W$-boson is taken to be $m_W = 80.419~{\rm
  GeV}$, the width $\Gamma_W = 2.140$~{\rm GeV}. $W$ couplings to
fermions are obtained from $\alpha_{\rm QED} (m_Z) = 1 / 128.802$ and
$\sin^2 \theta_W = 0.2222$.  We use MSTW08LO parton distribution
functions for leading order and MSTW08NLO for next-to-leading order
computations, corresponding to $\alpha_s(M_Z) = 0.13939$ and
$\alpha_s(M_Z) = 0.12018$ respectively~\cite{Martin:2009iq}. We do not
impose lepton isolation cuts. All results discussed below apply to the
QCD production $pp \to W^+ W^+ jj$; the electroweak contribution to
this process is ignored.  Note, however, that at next-to-leading
order, QCD and electroweak production processes start to interfere and
the separation of the two production mechanisms is not as clean as it
is at leading order.

We point out that, because a massive gauge boson is produced from each fermion line, the
cross-section for the process $pp \to W^+ W^+ jj$ remains finite {\it
  even} if the requirement that two jets are observed is
lifted. Hence, we can consider the production of same-sign gauge
bosons in association with $n$ jets $pp \to W^+ W^+ + n~{\rm jets}$,
where $n = 0, 1, 2$ or $n \ge 2$. It is useful to consider different
jet multiplicities separately because, depending on the number of
jets, backgrounds to $pp \to W^+ W^+ + n~{\rm jets}$ change.  Also, if
we are interested in $pp \to W^+ W^+ + n~{\rm jets}$ as a potential
background to New Physics processes, it is helpful to know how jetty
this process is.

\begin{figure}[t]
\begin{center}
\includegraphics[angle=0,scale=0.51]{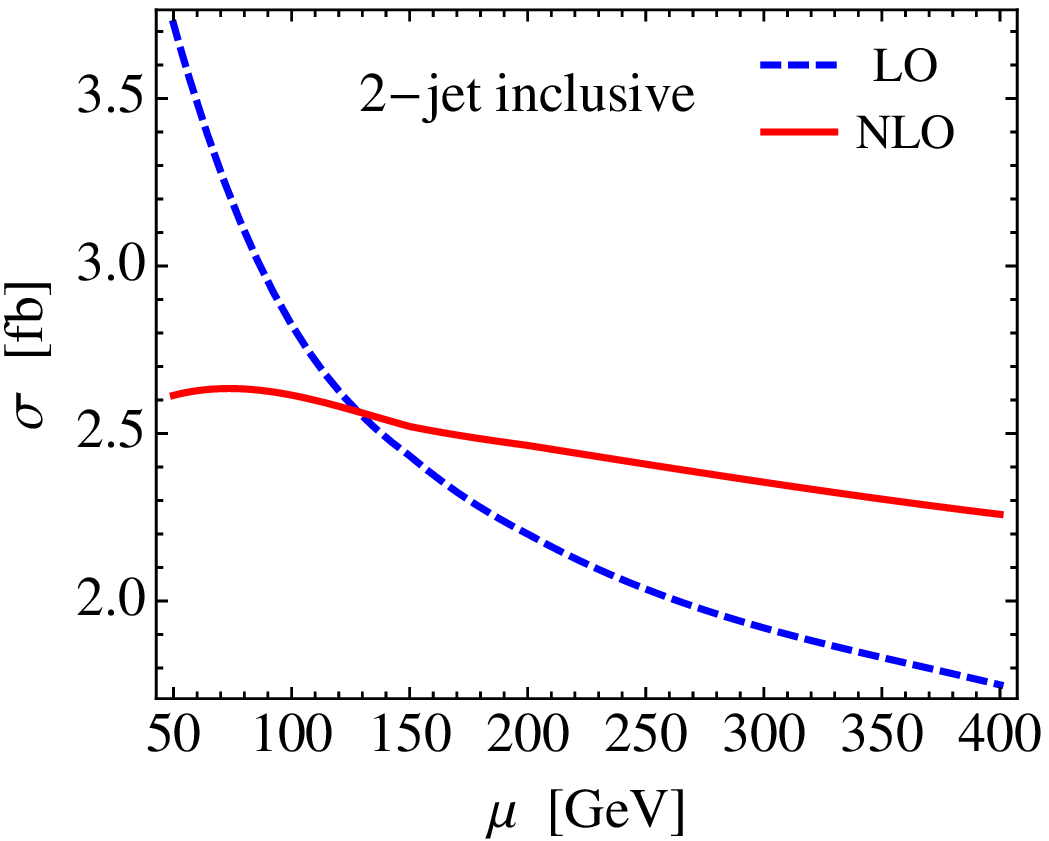}
\includegraphics[angle=0,scale=0.49]{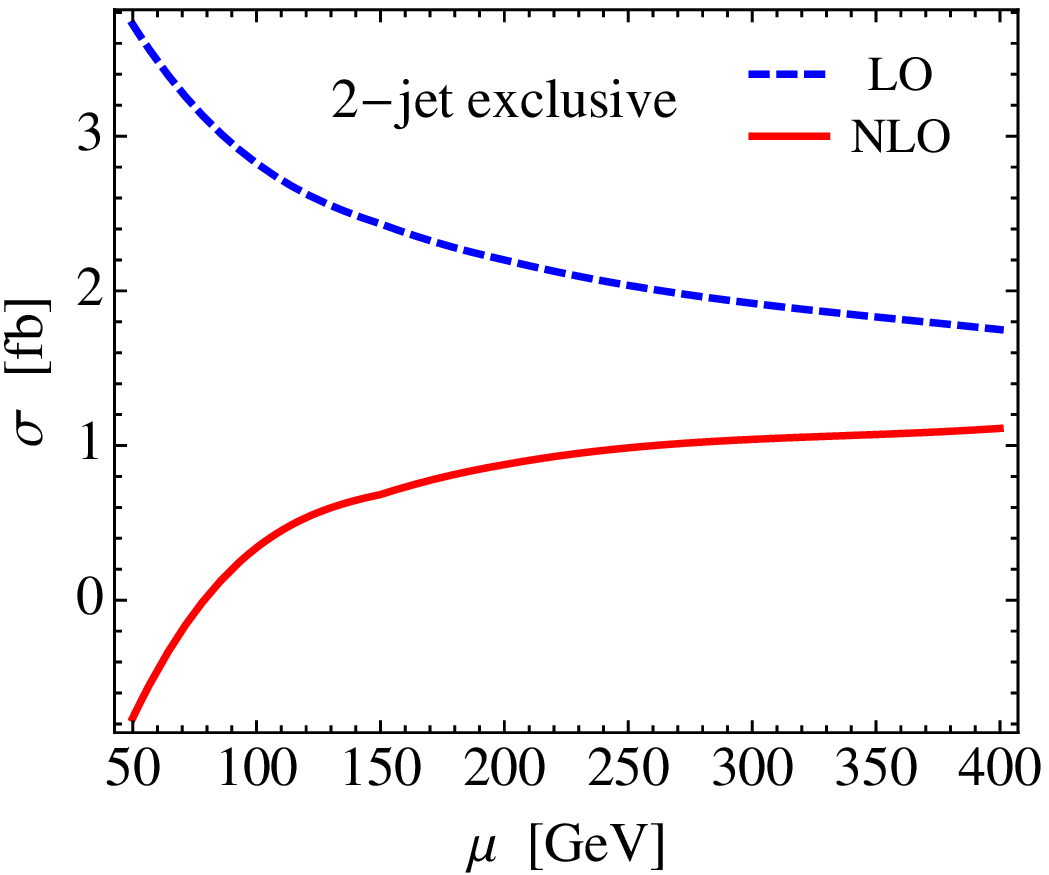}

\vspace{0.4cm}

\includegraphics[angle=0,scale=0.5]{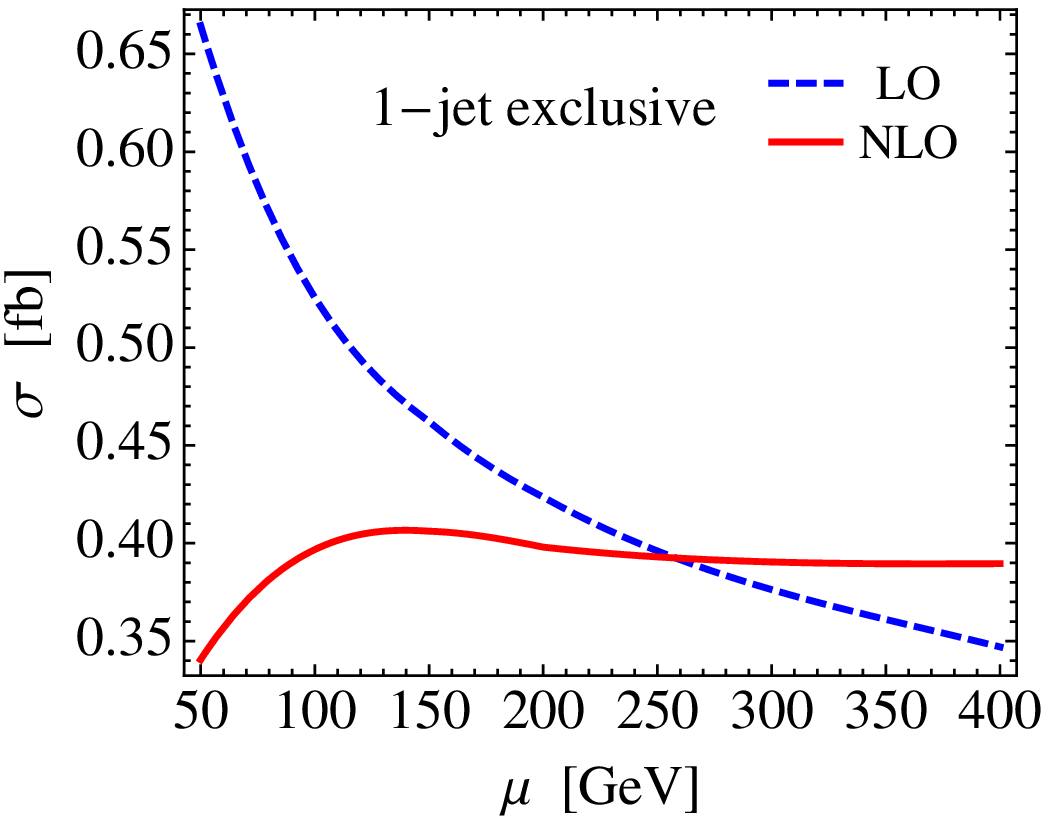}
\includegraphics[angle=0,scale=0.5]{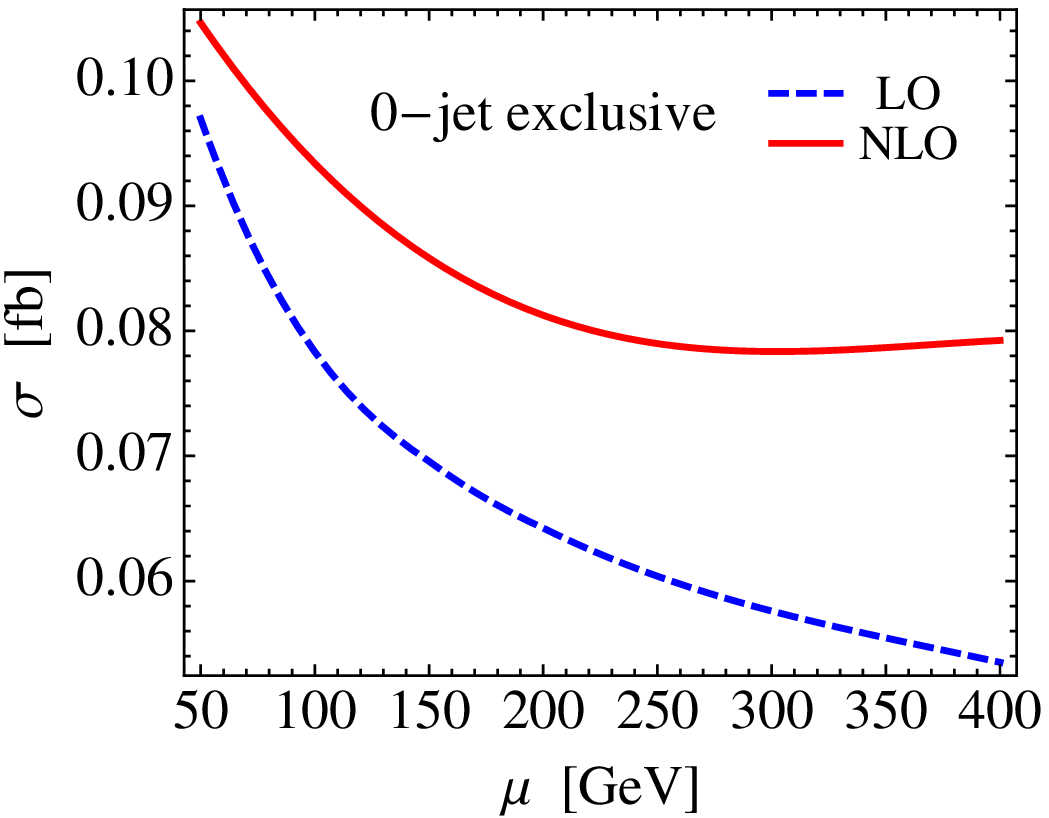}
\caption{The dependence on factorization and renormalization scales of
  cross-sections for $pp \to e^+\, \mu^+\, {\nu}_{e}\, {\nu}_{\mu} + n~{\rm
    jets}$, $n = 0,1,2$ at leading and next-to-leading order in
  perturbative QCD We set the two scales equal to each other $\mu_{\rm
    F} = \mu_{\rm R} = \mu$. }
\label{fig1}
\end{center}
\end{figure}

We begin by showing in Fig.~\ref{fig1} the dependence of the
production cross-sections for $pp \to e^+ \mu^+ \nu_{e} \nu_{\mu} +
n~{\rm jets}$ on the renormalization and factorization scales, which
we set equal to each other.
We show results for the following four processes
{\it i}) $pp \to W^+ W^+ + \ge 2~{\rm jets}$;
{\it ii}) $pp \to W^+ W^+ + 2~{\rm jets}$;
{\it iii})  $pp \to W^+ W^+ + 1~{\rm jet}$; and {\it iv}) 
$pp \to W^+ W^+ + 0~{\rm jets}$. 
The total inclusive cross-section is then given by the sum of the
two-jet inclusive cross-section, the one-jet exclusive cross-section
and the zero-jet exclusive cross-section. The difference between
inclusive and exclusive two jet cross-sections shows how often
$W^+W^+$ is produced in association with three, rather than two, jets.

It is clear from Fig.~\ref{fig1} that leading order cross-sections for
all jet multiplicities monotonically increase with the decrease of the
renormalization/factorization scales. This behavior is driven by the
dependence of the leading order cross-section on the square of the
strong coupling constant; the factorization scale dependence is
minor. It follows from Fig.~\ref{fig1} that when cross-sections are
calculated at next-to-leading order in perturbative QCD, they show
significantly reduced scale dependence.  For example, considering the
range of scales $50~{\rm GeV} \le \mu \le 400~{\rm GeV}$, we find the
two-jet inclusive cross-section to be $\sigma^{\rm LO} = 2.7 \pm
1.0~{\rm fb}$ at leading order and $\sigma^{\rm NLO} = 2.44 \pm
0.18~{\rm fb}$ at next-to-leading order. The forty percent scale
uncertainty at leading order is reduced to less than ten percent at
NLO.  We observe similar stabilization of the scale dependence for the
$0$- and $1$-jet exclusive multiplicities.  Combining these
cross-sections we obtain a total NLO cross-section of about $2.90~{\rm
  fb}$ for $pp \to e^+ \mu^+ \nu_e \nu_\mu $ inclusive
production. This implies about $60$ $e^+\mu^++ \e^+e^+ + \mu^+\mu^+$
events per year at the LHC with $10~{\rm fb}^{-1}$ annual
luminosity. While this is not a gigantic number, such events will have
a very distinct signature, so they will definitely be seen and it will
be possible to study them.

\begin{figure}[t]
\begin{center}
\includegraphics[angle=0,scale=0.7]{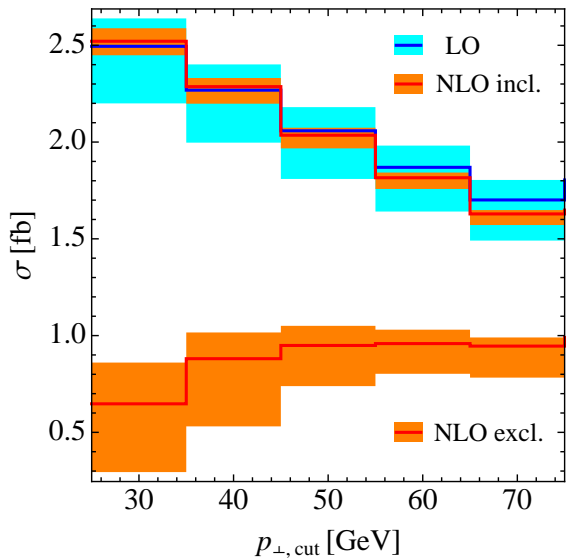}
\caption{The dependence on the jet $p_\perp$ cut of the two-jet
  inclusive and two-jet exclusive cross-sections for $pp \to e^+\,
  \mu^+\, {\nu}_{e}\, {\nu}_{\mu} + 2~{\rm jets}$ at leading and 
  next-to-leading
  order in perturbative QCD.  We show scale uncertainty bands, for
  values of the renormalization and factorization scales set to a
  common value $\mu$, which is varied in the interval $ 100~{\rm GeV}
  \le \mu \le 200~{\rm GeV}$.  Results for $\mu = 140~{\rm GeV}$ are
  shown as solid lines.}
\label{fig15}
\end{center}
\end{figure}

On the other hand, it is apparent from Fig.~\ref{fig1} that there is
quite a dramatic change in the two-jet {\it exclusive} cross-section.
At leading order, there is no difference between the exclusive and
inclusive cross-sections, but this difference appears at NLO. With
$\Delta R_{jj} = 0.4$ and a transverse momentum jet cut of $30~{\rm
  GeV}$, the two-jet exclusive cross-section is only about $0.7~{\rm
  fb}$ at $\mu = 150~{\rm GeV}$; if the scale is decreased to about
$80~{\rm GeV}$, the two-jet exclusive cross-section becomes negative.
One can argue that this is the consequence of the fact that the
$30~{\rm GeV}$ jet transverse momentum cut is too small for the
convergence of the perturbative expansion of the two-jet exclusive
cross-section. However, it is not fully clear how to make this
explanation compatible with the reasonable perturbative stability of
the zero-jet and the one-jet exclusive cross-sections for the same
value of the jet transverse momentum cut.  In Fig.~\ref{fig15} we show
the inclusive and exclusive cross-sections for $pp \to e^+ \mu^+
{\nu}_{e} {\nu}_{\mu} + 2~{\rm jets}$ at next-to-leading order in
perturbative QCD, in dependence on the jet $p_\perp$ cut.  We set the
factorization and renormalization scales equal to each other $\mu_R =
\mu_F = \mu$, and vary the scale $\mu$ in the range $100~{\rm GeV}
\leq \mu \leq 200$ GeV. It is clear that for a jet cut of $40-50~{\rm
  GeV}$, the scale dependence of the exclusive two-jet cross-section
is relatively small, while for the $30~{\rm GeV}$ jet cut, the scale
dependence increases strongly. The results displayed in
Fig.~\ref{fig15} suggest that, whatever the exact value of the
exclusive two-jet cross-section is, it is significantly smaller than
the two-jet inclusive cross-section. This smallness implies that quite
a large fraction of events in $pp \to e^+\mu^+ \nu_e \nu_\mu + \ge
2~{\rm jets}$ have a relatively hard third jet. This effect remains present, although less pronounced, for larger values of $R$.  This feature may be
useful for rejecting contributions of $pp \to W^+ W^+ jj$ when looking
for multiple parton scattering.

We now turn to the discussion of kinematic distributions.  Unless
explicitly stated otherwise, we will consider two-jet inclusive
processes.  A glance at Fig.~\ref{fig1} suggests that NLO QCD
corrections to jet cross-sections are small if the renormalization and
factorization scales are set to $\mu = 150~{\rm GeV}$.  
Indeed, for $\mu = 150~{\rm GeV}$, two-jet inclusive leading and
next-to-leading order cross-sections nearly coincide $\sigma_{\rm LO}
= 2.4~{\rm fb}$ and $\sigma_{\rm NLO} = 2.5~{\rm fb}$, so that the
NLO QCD corrections change the leading order cross-sections by less
than five percent.
In the plots that follow, we show the scale uncertainty bands for
$50~{\rm GeV} \leq \mu \leq 400~{\rm GeV}$ and predictions for $\mu =
150~{\rm GeV}$, for leading and next-to-leading order distributions.

\begin{figure}[t]
\begin{center}
\includegraphics[angle=0,scale=0.51]{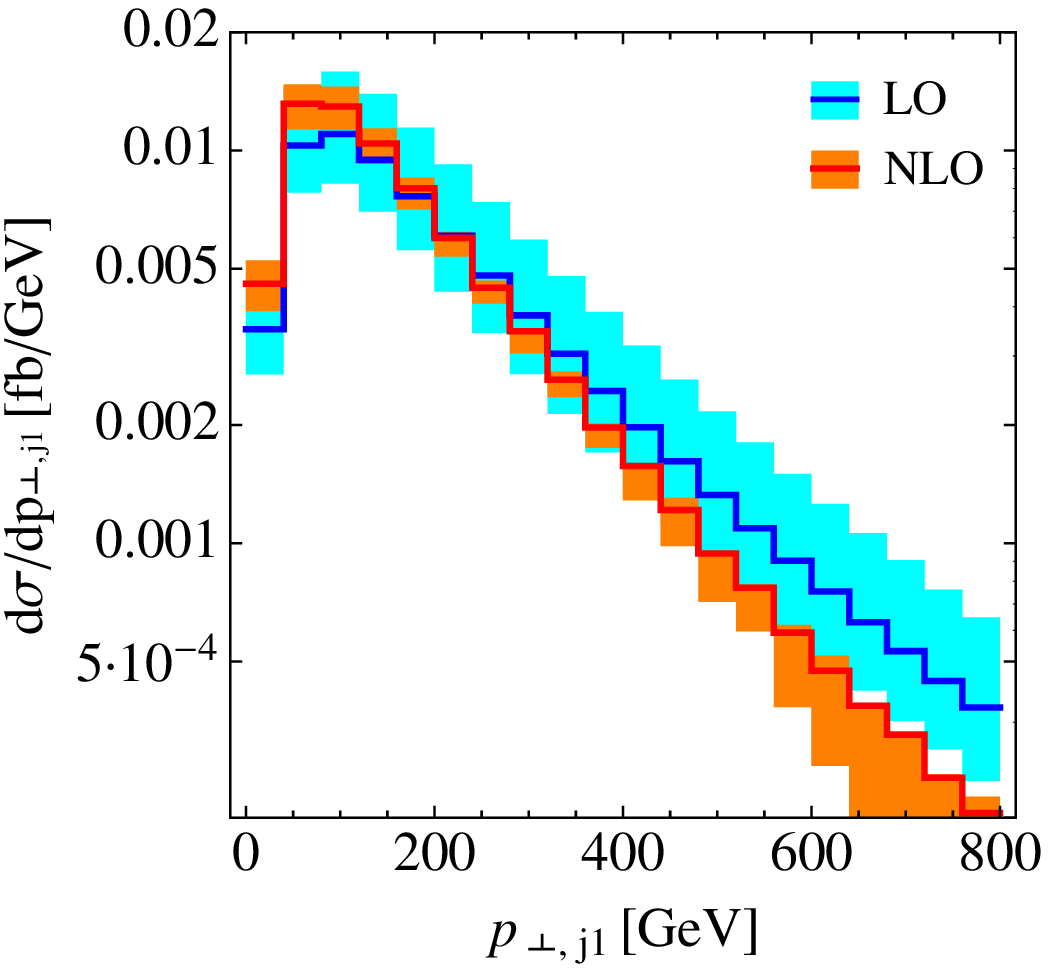}
\qquad
\includegraphics[angle=0,scale=0.49]{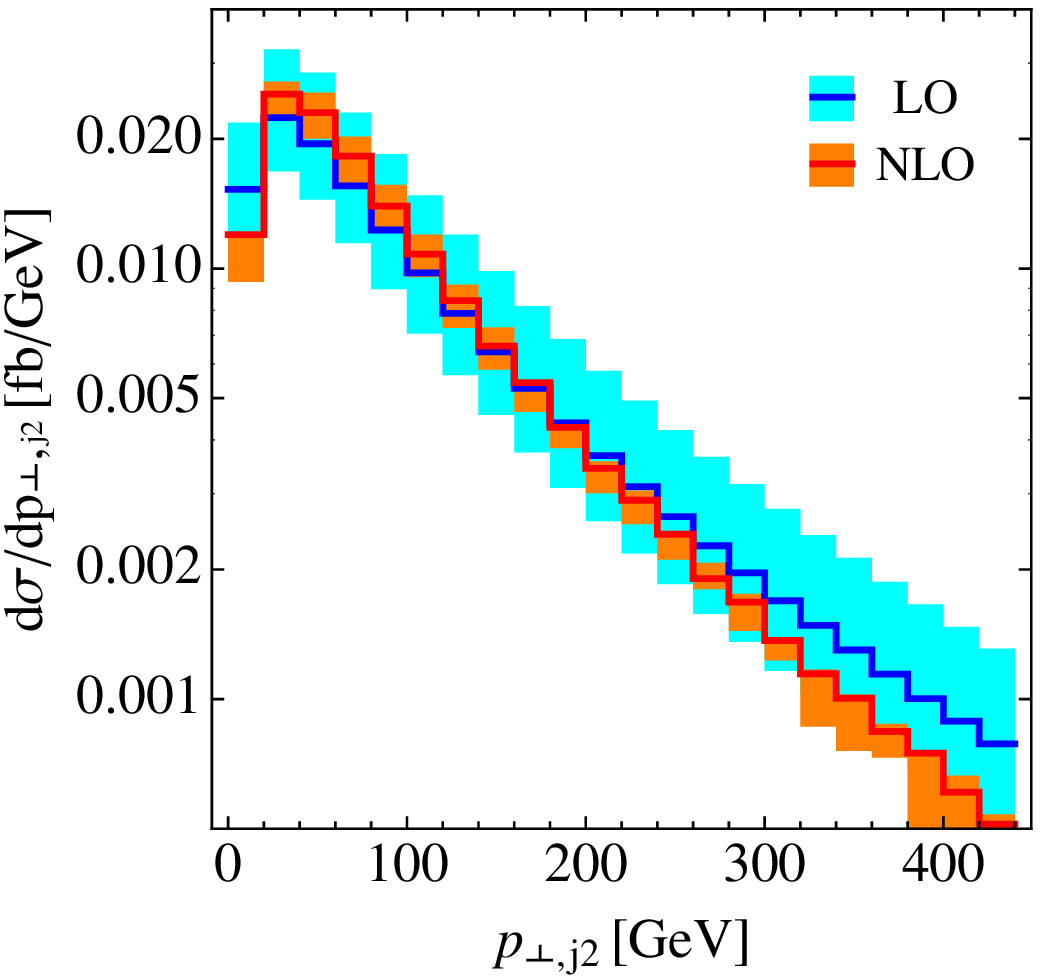}

\vspace{0.2cm}

\includegraphics[angle=0,scale=0.53]{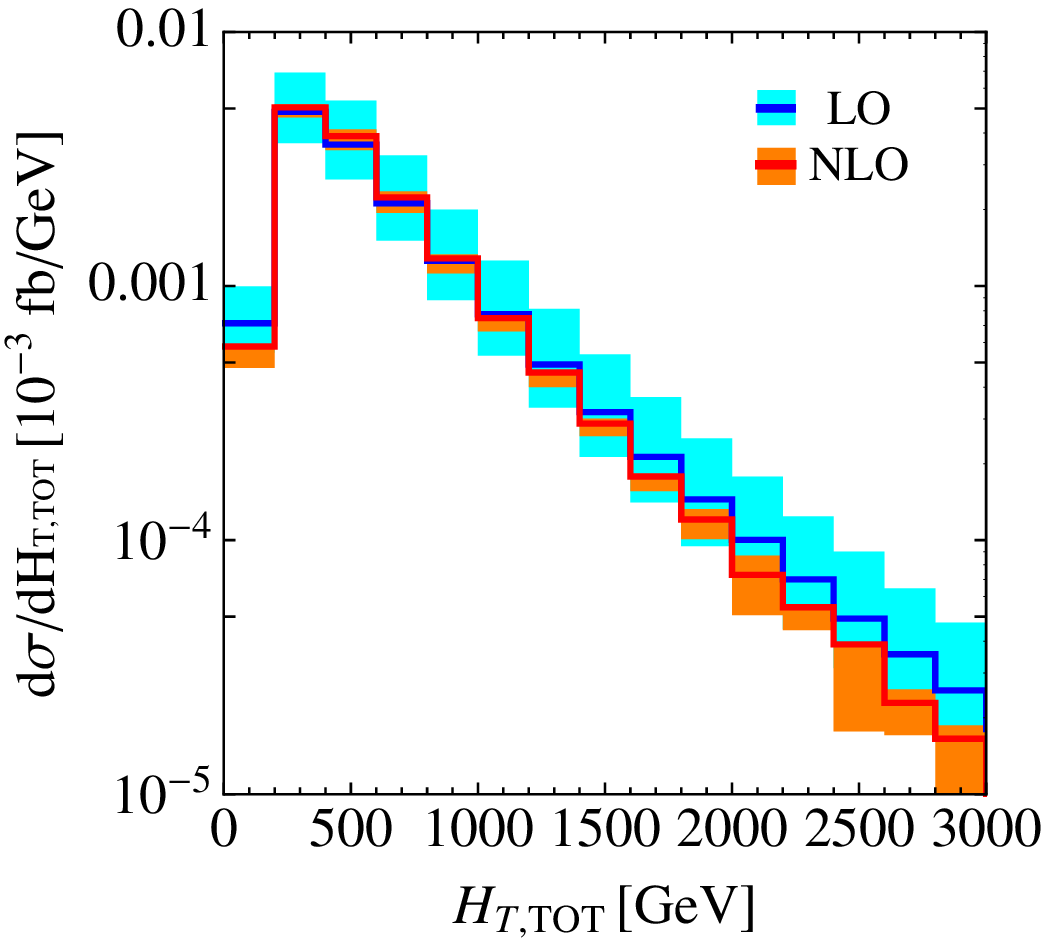}
\qquad
\includegraphics[angle=0,scale=0.47]{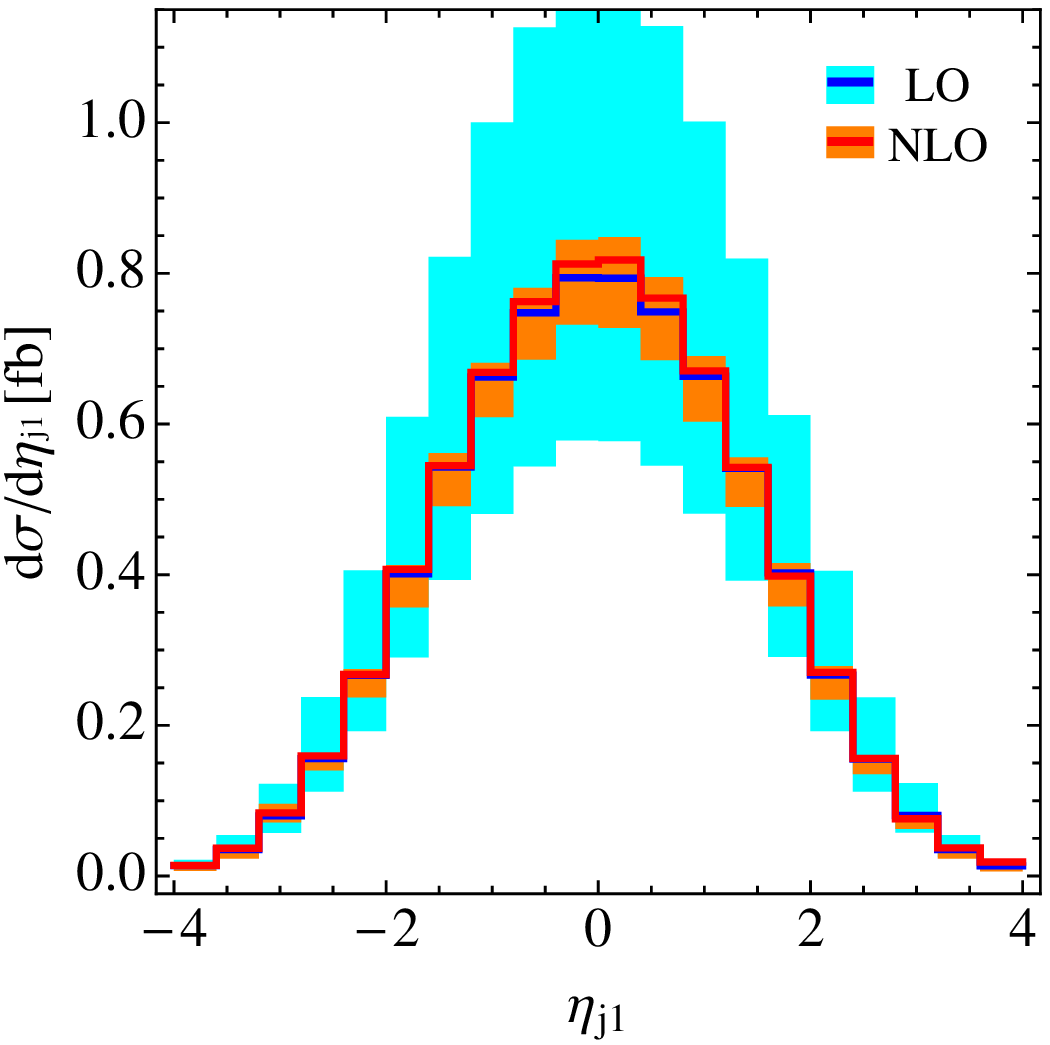}

\vspace{0.2cm}

\includegraphics[angle=0,scale=0.49]{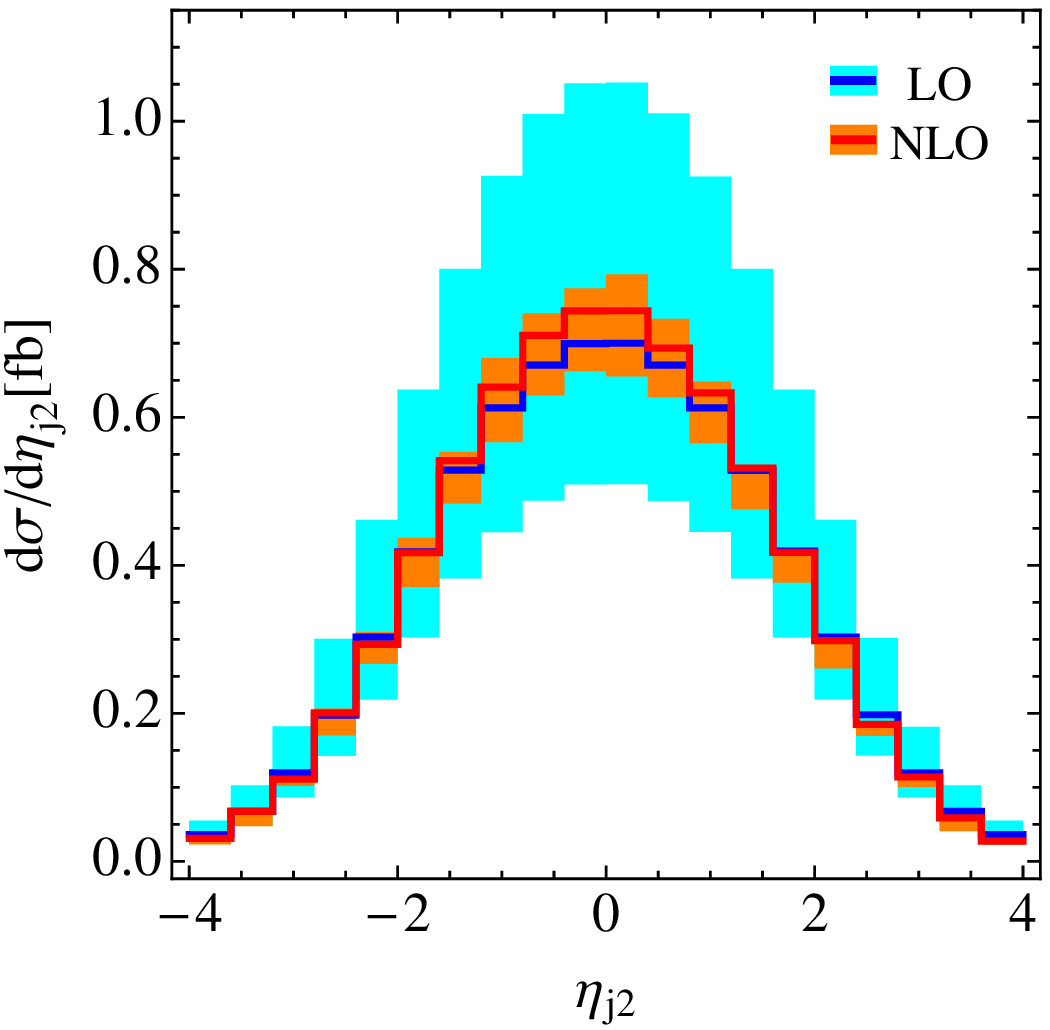}
\qquad
\includegraphics[angle=0,scale=0.49]{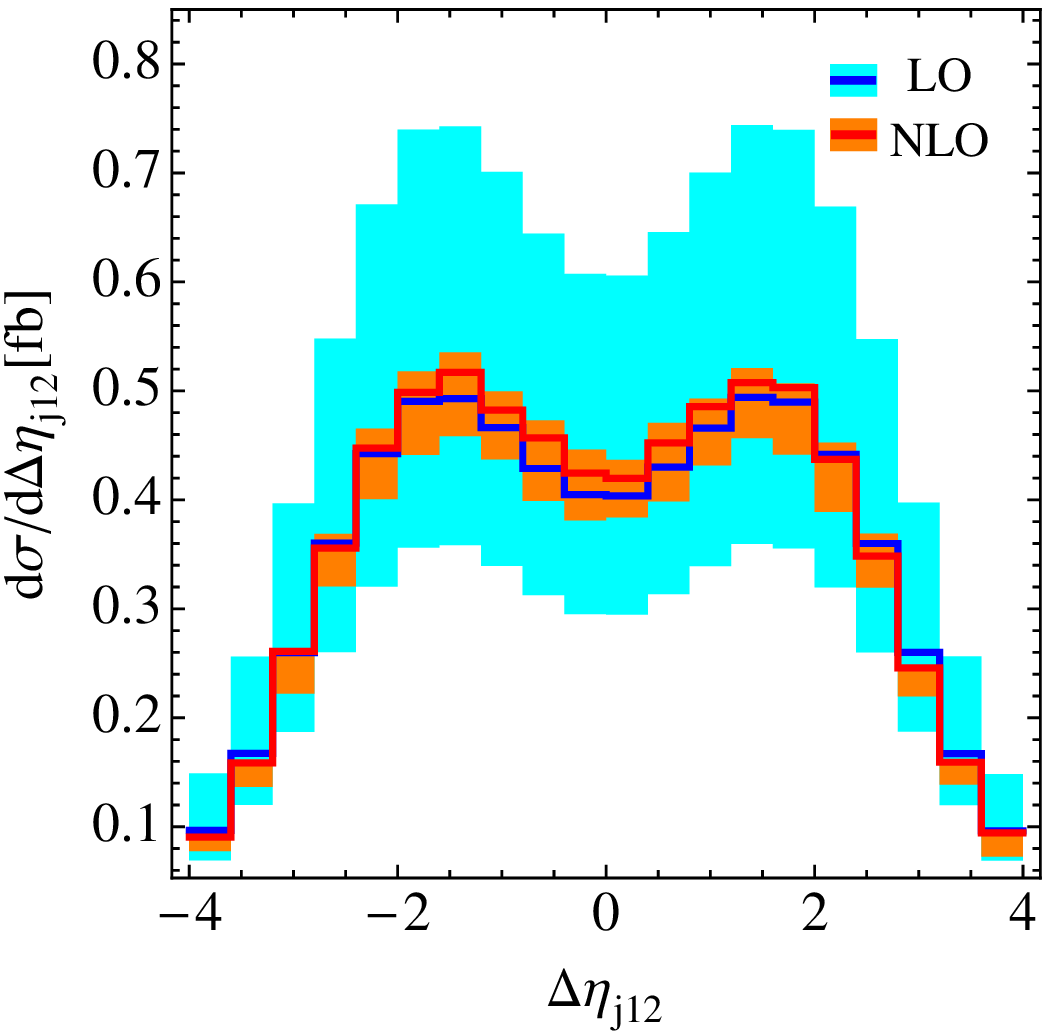}
\caption{Kinematic distributions of the two hardest jets in the
  process $pp \to e^+\, \mu^+\, {\nu}_{e}\, {\nu}_{\mu}\, + 2~{\rm jets}$ at
  leading and next-to-leading order in perturbative QCD for inclusive
  two-jet events.  The bands show renormalization and factorization
  scale uncertainty, for $50~{\rm GeV} \leq \mu \leq 400~{\rm
    GeV}$. Solid lines show leading and next-to-leading order
  predictions for $\mu = 150~{\rm GeV}$.  We use $H_{\rm T,TOT} = \sum
  \limits_{j}^{} p_{\perp,j} + p_\perp^{e^+} + p_\perp^{\mu^+} +
  p_\perp^{\rm miss}$.  }
\label{fig2}
\end{center}
\end{figure}

The transverse momentum distributions of the hardest and
next-to-hardest jets, as well as the distribution of the scalar sum of
the transverse momenta of all jets and charged leptons and the missing
transverse momentum in the event are shown in the first three pane of
Fig.~\ref{fig2}. It is clear from Fig.~\ref{fig2} that jets in $pp \to
W^+ W^+ jj$ are hard; a typical transverse momentum of the hardest jet
is close to $100~{\rm GeV}$ and the transverse momentum of the
next-to-hardest jet is close to $40~{\rm GeV}$.  While no major
changes in kinematic distributions occur when NLO QCD corrections are
calculated, some shape changes are apparent from Fig.~\ref{fig2}.
Indeed, the NLO QCD transverse momenta distributions of the hardest
and next-to-hardest jet show a characteristic depletion at large
values of $p_{\perp,j}$.  One reason this change occurs is because a
constant, rather than a dynamical, renormalization scale is used in
our leading order calculation. As was emphasized many times recently,
the choice of dynamical renormalization scales in leading order
computations may affect shapes of kinematic distributions in a manner
similar to NLO QCD corrections
\cite{jager,Berger:2009ep,Melnikov:2009wh,nlowork}.  The shapes of
rapidity distributions of the hardest and next-to-hardest jets, as
well as the distribution of the rapidity differences between hardest
and next-to-hardest jets, shown in the next three pane, remain very
similar to the shapes of leading order distributions, once the NLO QCD
corrections are included.  On the other hand, scale dependencies of
rapidity distributions are reduced dramatically.

\begin{figure}[t]
\begin{center}
\includegraphics[angle=0,scale=0.51]{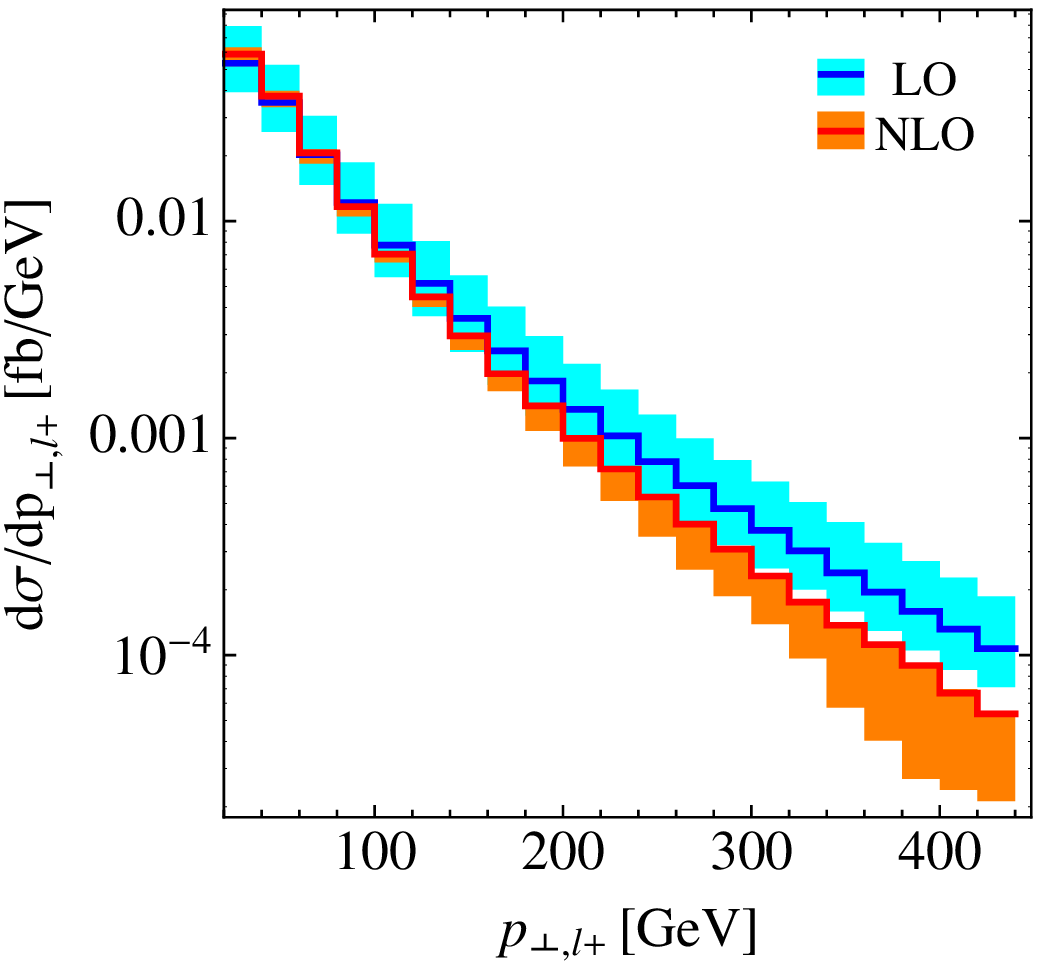}
\qquad
\includegraphics[angle=0,scale=0.52]{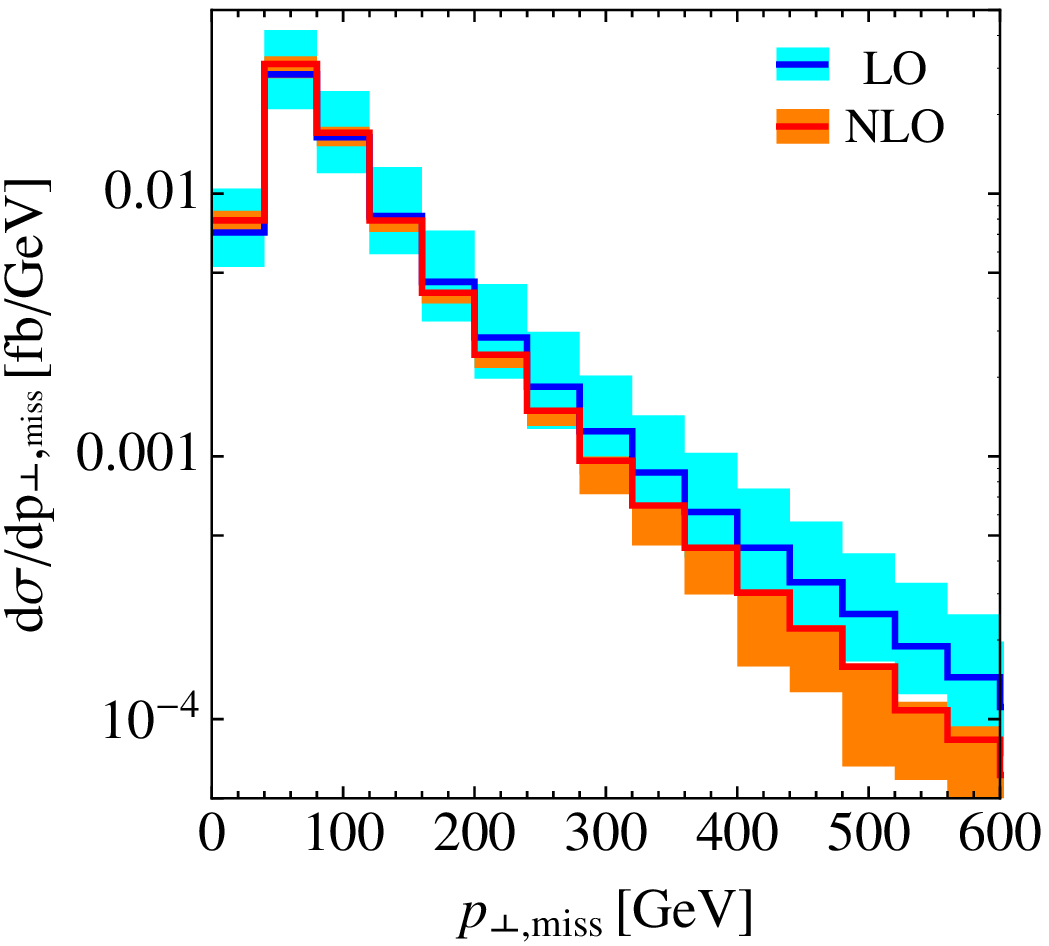}

\vspace{0.2cm}

\includegraphics[angle=0,scale=0.5]{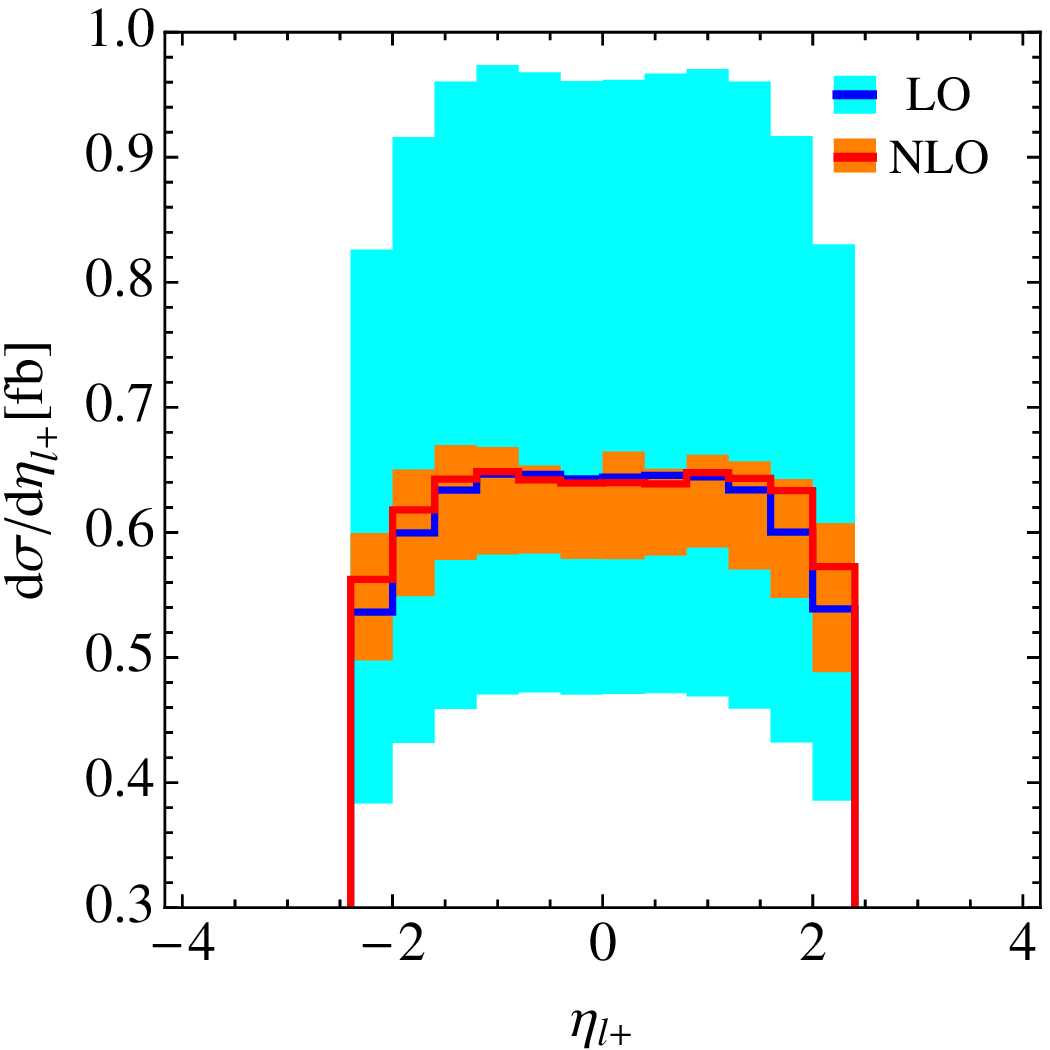}
\qquad
\includegraphics[angle=0,scale=0.5]{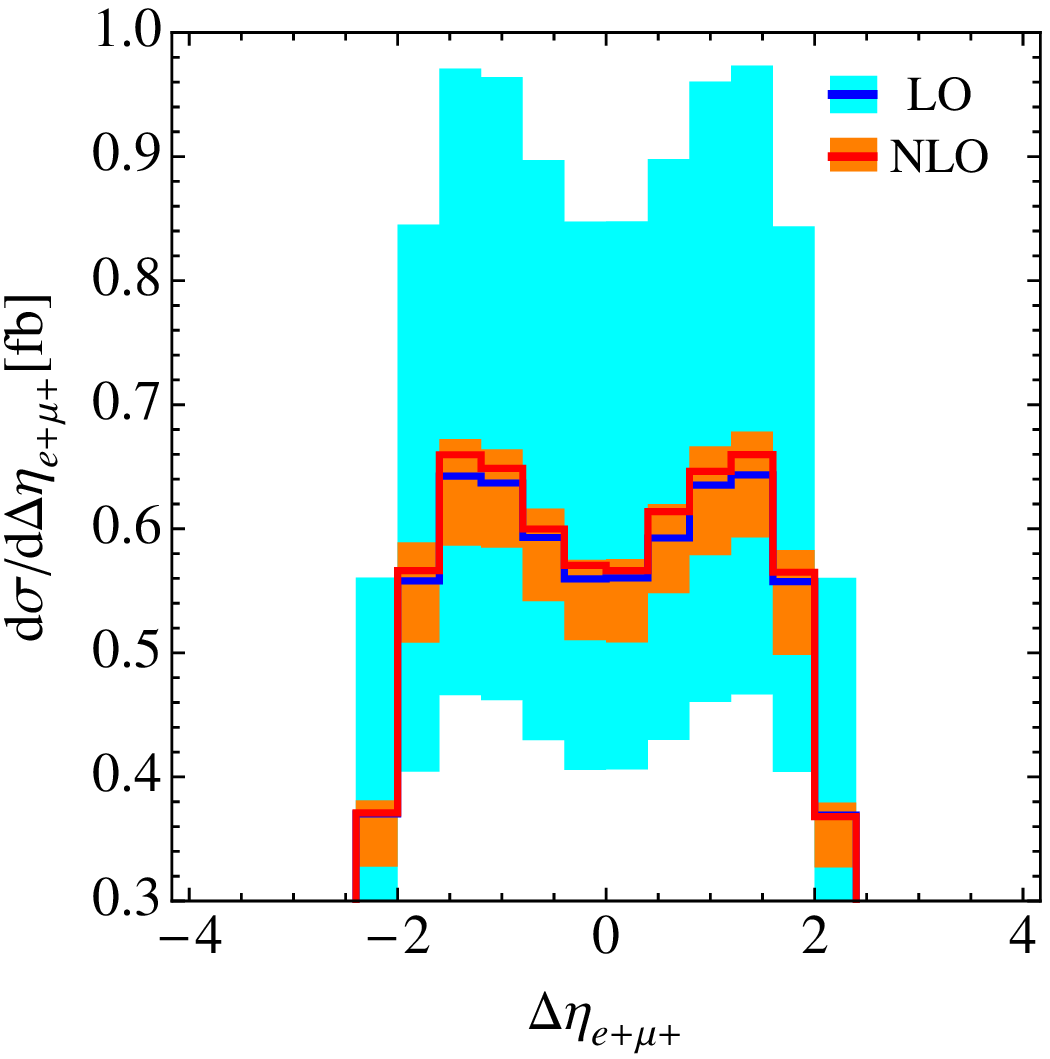}

\vspace{0.2cm}

\includegraphics[angle=0,scale=0.54]{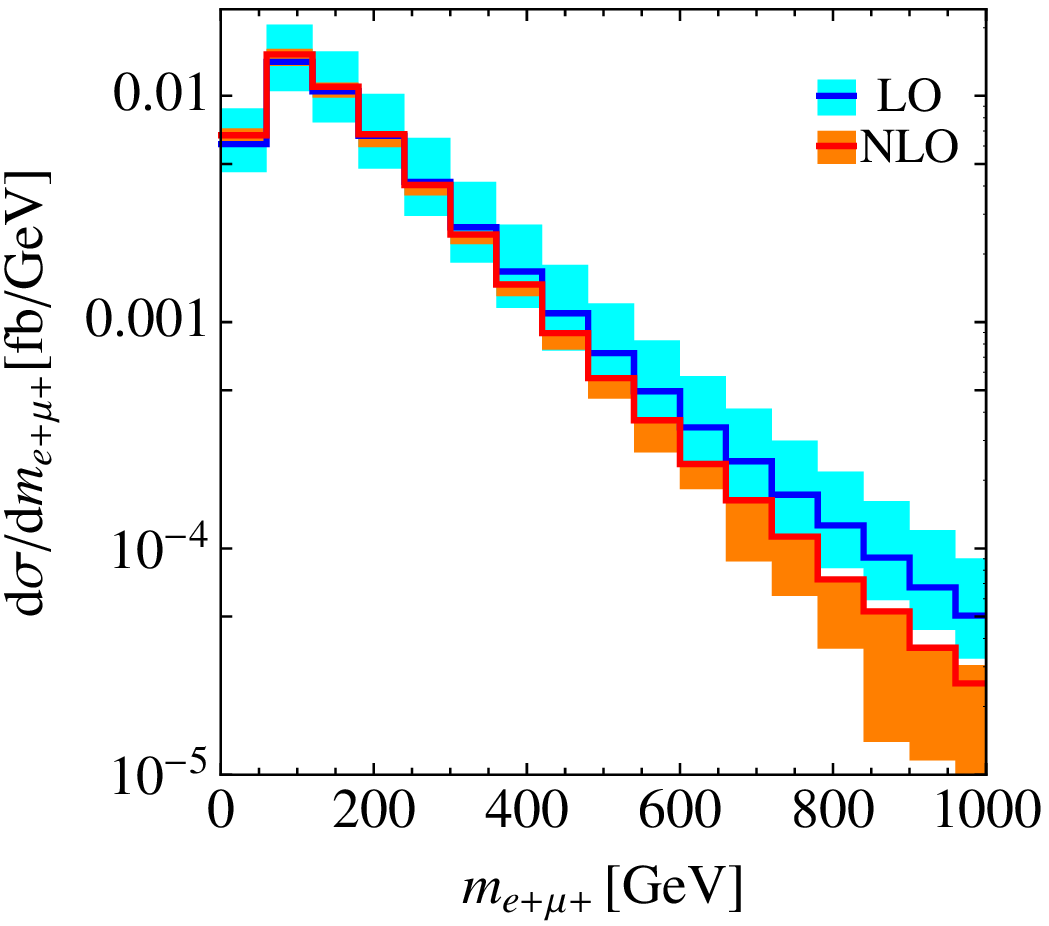}
\quad
\includegraphics[angle=0,scale=0.52]{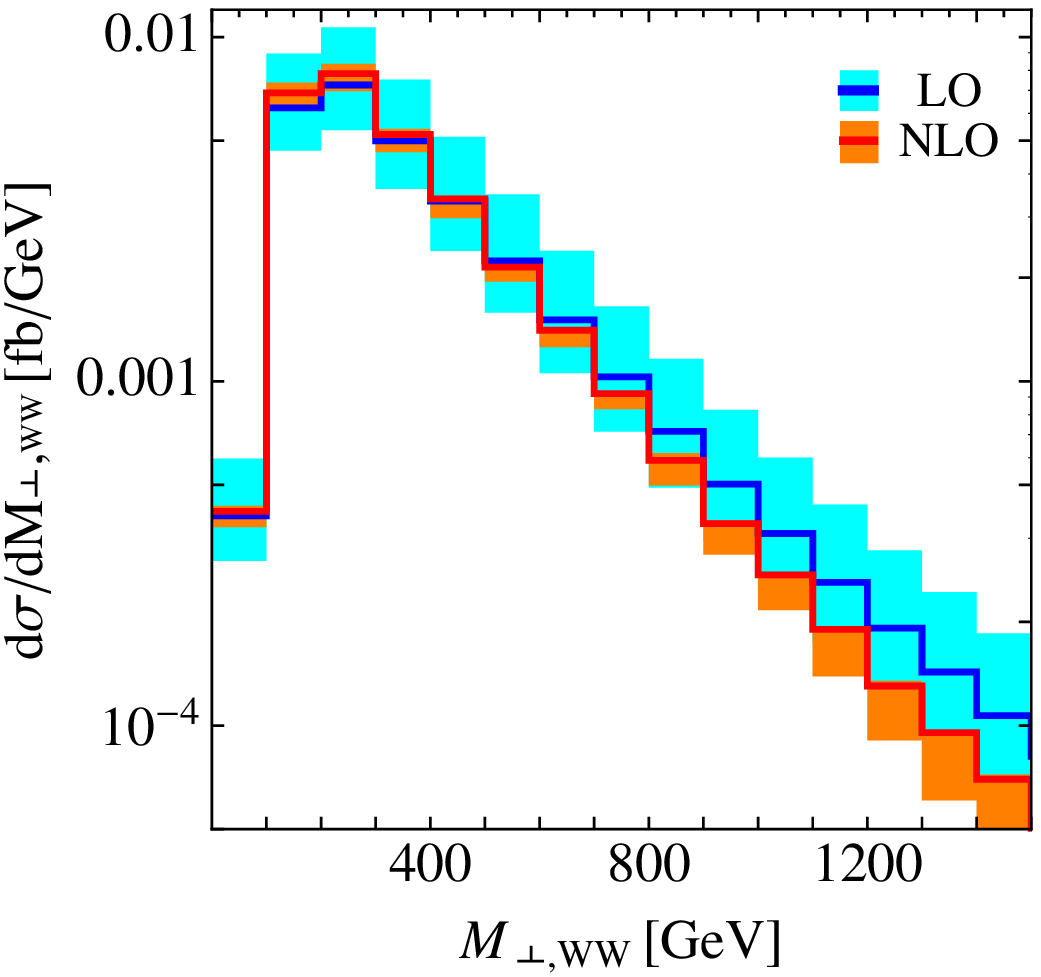}
\caption{Kinematic distributions of charged leptons and missing energy
  in $pp \to e^+\, \mu^+\, {\nu}_{e}\, {\nu}_{\mu}\, + n~{\rm jets}$ at
  leading and next-to-leading order in perturbative QCD for inclusive
  two-jet events.  The bands show renormalization and factorization
  scale uncertainty, for $50~{\rm GeV} \leq \mu \leq 400~{\rm
    GeV}$. Solid lines show leading and next-to-leading order
  predictions for $\mu = 150~{\rm GeV}$.  }
\label{fig3}
\end{center}
\end{figure}

\begin{figure}[t]
\begin{center}
\includegraphics[angle=0,scale=0.5]{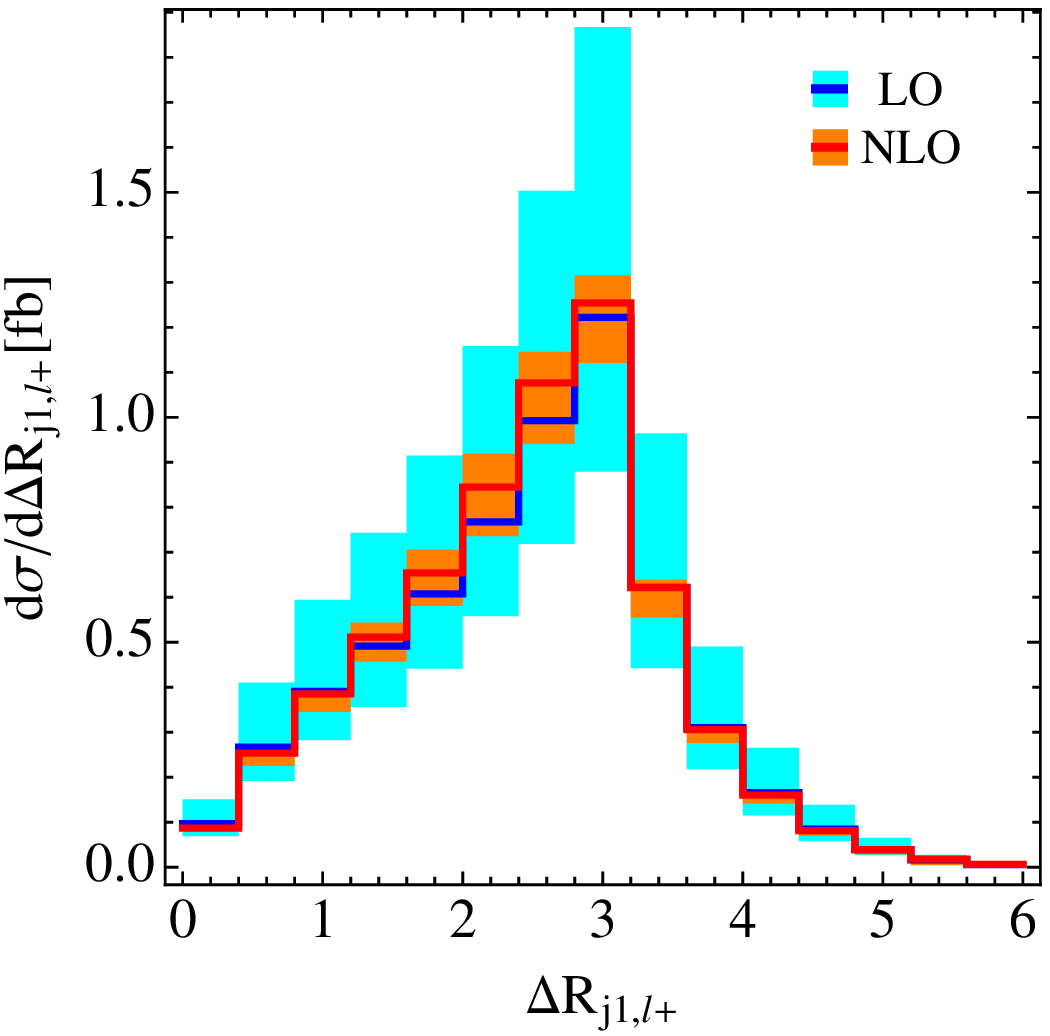}
\qquad 
\includegraphics[angle=0,scale=0.5]{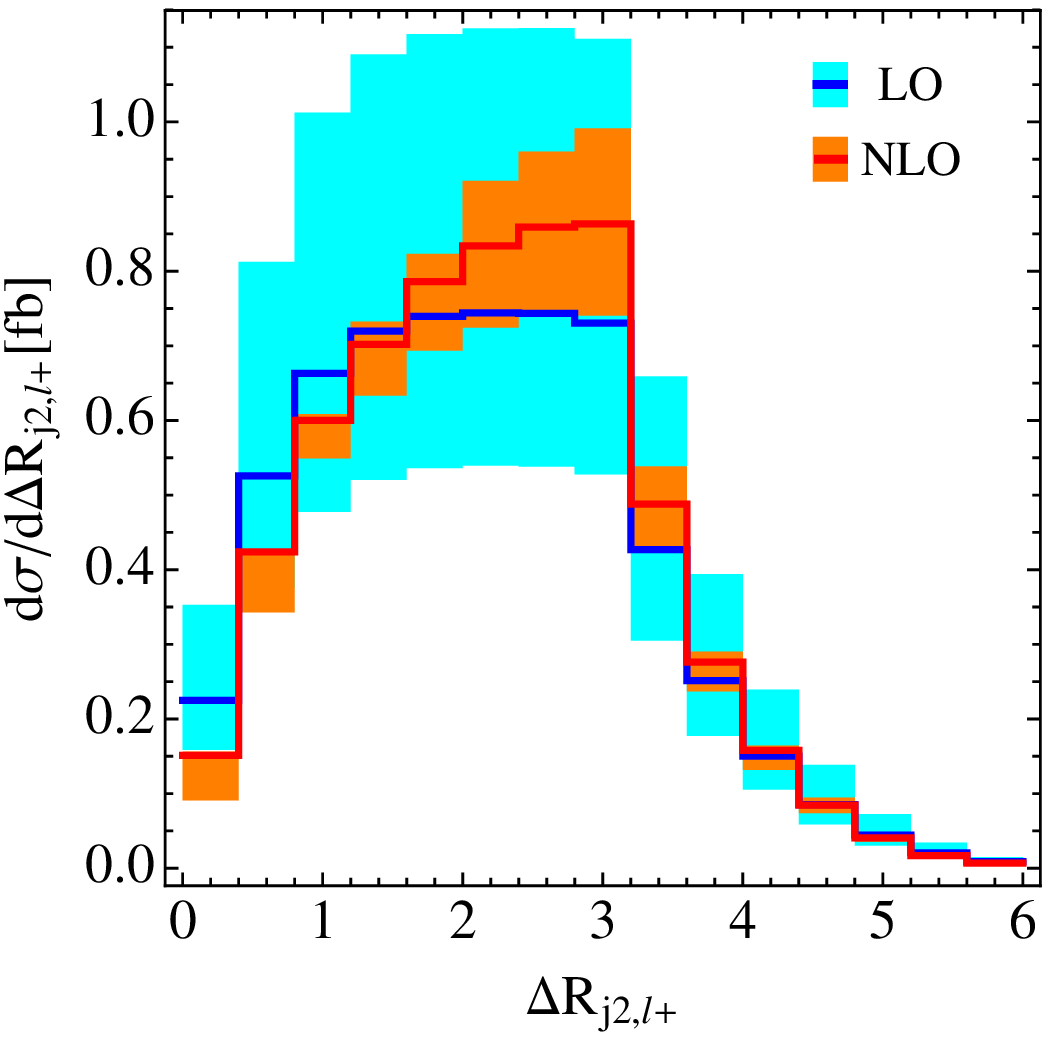}

\vspace{0.2cm}

\includegraphics[angle=0,scale=0.5]{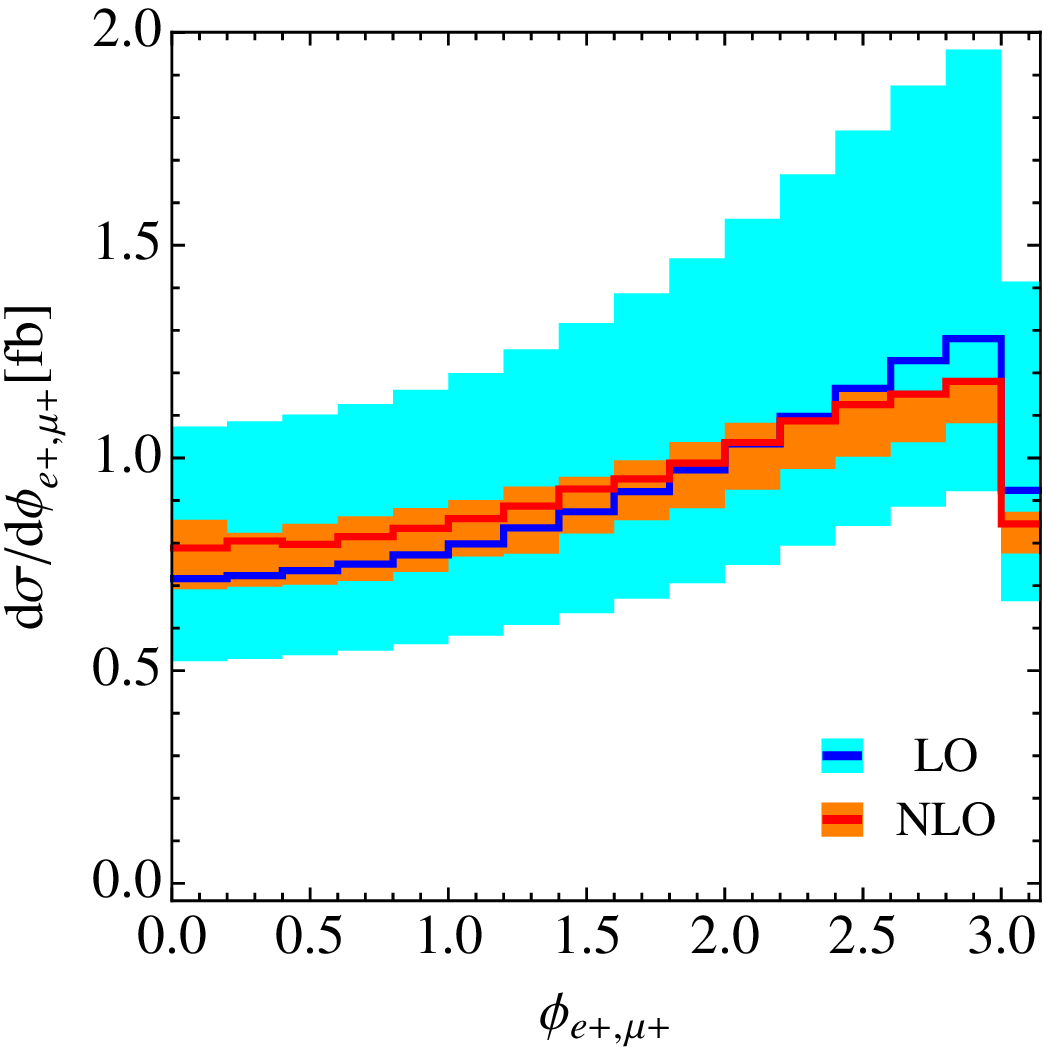}
\caption{Angular kinematic distributions  in the process 
$pp \to e^+ \mu^+
  \nu_{e} \nu_{\mu} + 2~{\rm jets}$ at leading and next-to-leading
  order in perturbative QCD for inclusive two-jet events.  The bands
  show renormalization and factorization scale uncertainty, for
  $50~{\rm GeV} \leq \mu \leq  400~{\rm GeV}$. Solid lines show leading and
  next-to-leading order predictions for $\mu = 150~{\rm GeV}$.  }
\label{fig4}
\end{center}
\end{figure}

We now turn to lepton and missing transverse momentum
distributions. In Fig.~\ref{fig3} we show the charged lepton
transverse momentum distribution, the missing transverse momentum
distribution, the distributions of the charged lepton rapidity and of
the rapidity difference of the two leptons, as well as the
distributions of the invariant mass of the two charged leptons and the
transverse mass of the two $W$-bosons \footnote{Since neutrinos are
  not observed, it is impossible to reconstruct the true transverse
  mass of the pair of the two $W$-bosons. We follow Ref.\cite{eboli}
  and define $m_{\rm {\perp},WW}^2 = (E_{\perp, l^+ l^+} +
  \tilde{E}_{\perp, \rm miss})^2 - ( {\bf p}_{\perp, l^+ l^+} + {\bf
    p}_{\perp, \rm miss})^2$, where the missing transverse energy
  $\tilde{E}_{\perp, \rm miss}$ is reconstructed from the missing
  transverse momentum using the invariant mass of the charged lepton
  system $\tilde{E}_{\perp, \rm miss} = \sqrt{{\bf p}_{\perp, \rm
      miss}^2 + m_{l^+ l^-}^2}$.}. Lepton kinematic distributions are
affected by the QCD radiative corrections in a similar way to the jet
kinematic distributions. We observe that transverse momentum
distributions become softer while shapes of rapidity distributions are
unaffected.  Distributions of dilepton invariant mass and the
transverse mass of the $W$-bosons become softer as well.  It is also
interesting to look at the angular distributions of the charged
leptons. In Fig.~\ref{fig4} the angular distance $\Delta R_{lj} =
\sqrt{(\eta_l - \eta_j)^2 + (\phi_l - \phi_j)^2}$ between a charged
lepton of fixed flavor ($e^+$ or $\mu^+$) and the hardest
(next-to-hardest) jet is displayed, as well as the distribution of the
relative azimuthal angle of the two charged leptons. Although
distributions of angular distances between leptons and jets are broad,
they peak at $\Delta R_{lj} \approx 3$ for both hardest and
next-to-hardest jets.  NLO QCD effects do not change this conclusion
but, interestingly, they make the angular distance between
next-to-hardest jet and the charged lepton somewhat {\it larger} at
next-to-leading order. The distribution of the relative azimuthal
angle of the two charged leptons becomes less peaked at $\Delta
\phi_{l^+ l^+} = \pi$, although the two leptons still prefer to be
back to back.
It is interesting to remark that, if the two same sign leptons are
produced through a double-parton scattering mechanism, their
directions are not correlated. Hence, yet another possibility to
reduce the single-scattering-background is to cut on the relative
azimuthal angle between the two leptons.

\section{Conclusion}
\label{sc4}

In this paper, we presented the computation of NLO QCD corrections to
the QCD-mediated process $pp \to W^+ W^+ jj$ at the LHC. In spite of
the fact that this is a $2 \to 4$ process, the computation of NLO QCD
corrections is relatively straightforward, thanks to spectacular
developments in technology of NLO QCD computations in recent years
\cite{denner,golem,britto,Britto:2004tx,Forde:2007mi,opp,opp2,egk,Giele:2008ve,cfb}.
We note that NLO QCD corrections to electroweak-mediated $pp \to
W^+W^+jj$ process were recently calculated in Ref.~\cite{jager}.

The production of two same-sign $W$-bosons in association with two
jets is a background to studies of double-parton scattering, as well
as to a number of beyond the Standard Model physics processes.  The
NLO QCD corrections to $pp \to W^+ W^+ jj$ reduce the scale dependence
to about ten percent.  An interesting feature of $pp \to W^+ W^+ jj$
is that perturbative QCD is applicable even if the observation of the
two jets is not demanded. This opens up the possibility to study QCD
radiative corrections to $pp \to W^+ W^+ +n~{\rm jets}$ where the
number of jets can be zero, one and two. At next-to-leading order, the
final state with three jets also appears. We have studied QCD
corrections to processes with various jet multiplicities and observed
reasonable behavior of the perturbative expansion for two-jet
inclusive, and one- and zero-jet exclusive processes. The situation
with the two-jet exclusive cross-section is less satisfactory. As
follows from Fig.~\ref{fig15}, the two-jet exclusive cross-section is
about fifty percent of the inclusive cross-section. Since at leading
order exclusive and inclusive two-jet cross-sections coincide, the NLO
QCD correction is large.  In addition, there is significant residual
scale uncertainty in the prediction for the exclusive two-jet
cross-section which is smaller than, but comparable to, the leading
order scale dependence.
We believe that these results suggest that nearly fifty percent of all
events in $pp \to e^+ \mu^+ \nu_e \nu_\mu + \ge 2~{\rm jets}$ contain
a relatively hard third jet.  It may be possible to use this effect in designing
selection or suppression criteria for $W^+W^+jj$ final state. We have
also studied a variety of kinematic distributions and showed that no
dramatic shape changes occur.

As a final comment, we point out that the production of two {\it
  negatively} charged $W$-bosons at the LHC is also possible, at a
rate which is about forty percent of the $W^+W^+$ production
\cite{dps}.  We can use our set up to compute NLO QCD corrections to
$pp \to W^-W^-jj$ using charge conjugation and parity reversal
\cite{jager}. To this end, if we are interested in $pp \to e^-\mu^-
\bar \nu_e \bar \nu_\mu jj$, we can do a calculation for $\bar p \bar
p \to e^+ \mu^+ \nu_e \nu_\mu jj$, treat the final state leptons as if
they are positively charged and reverse the momentum directions for
parity-odd distributions.

{\bf Acknowledgments} 
We thank Anna Kulesza, Ezio Maina and Dieter Zeppenfeld for comments
on the relative magnitude of the QCD- and electroweak-mediated
contributions to $pp \to W^+W^+jj$. We acknowledge useful
conversations with Ed Berger, Zoltan Kunszt and Paolo Nason. We thank
Valentin Hirschi and Thomas Reiter for pointing out an error in the
Tables. During the work on this paper, we have benefited from the
hospitality extended to us by Aspen Center for Physics and CERN Theory
Division.  This research is supported by the NSF under grant
PHY-0855365, by the start up funds provided by Johns Hopkins
University and by the British Science and Technology Facilities
Council.

\newpage 
\appendix

\section{Results for primitive amplitudes at a fixed phase space point}

In this Appendix we give numerical results for all primitive
amplitudes needed to reconstruct the one-loop amplitudes in
Eq.~(\ref{eq1}) at a particular phase space point. The momenta
$(E,p_x,p_y,p_z)$, in GeV, are chosen to be
\begin{eqnarray}
&&k_1({\bar{u}}) = (-500.000000000000,-500.000000000000,0.00000000000000,0.00000000000000) ,\nn\\
&&k_2({d}) =  (-500.000000000000, 500.000000000000,0.00000000000000,0.00000000000000) , \nn\\
&&k_3({\bar{c}}) = ( 54.2314070117999,-31.1330162081798,-7.92796656791140,43.6912823611163) , \nn\\
&&k_4({s}) = ( 214.488870161418,-27.0607980217775,-98.5198083786150,188.592247959949) .\nn\\
&&k_5({e^+}) = ( 85.5312248384887,-8.22193223977868, 36.1637837682033,-77.0725048002414) , \nn\\ 
&&k_6(\nu_e) = ( 181.428811610043,-57.8599829481937,-171.863734086635,-5.61185898481311) ,  \nn\\
&&k_7({\mu^+}) = ( 82.8493010774356,-65.9095476235891,-49.8952157196287, 5.51413360058664) ,  \nn\\
&&k_8({\nu_\mu}) = ( 381.470385300815, 190.185277041519,292.042940984587,-155.113300136598) . \nn\\
\label{eq:point}
\end{eqnarray}

Because the $W$-bosons couple only to left-handed fermions and
right-handed anti-fermions, all helicities in the following are fixed,
so that $\{\bar{u},d,\bar{c},s,e^+,\nu_e,\mu^+,\nu_\mu\}$ have
helicities $\{+,-,+,-,+,-,+,-\}$. Below we tabulate numerical results
for the primitive tree and one-loop amplitudes at the phase-space
point given in Eq.~(\ref{eq:point}).  In Table~\ref{tab1} we display
tree-level amplitudes and ratios of {\it unrenormalized} one-loop
amplitudes to tree-amplitudes
\begin{equation}
r_i= \frac{1}{c_\Gamma}\frac{A_i}{A_0}
\end{equation} 
 where
$\displaystyle
c_\Gamma = \frac{\Gamma(1+\epsilon)\Gamma^2(1-\epsilon)}{(4\pi)^{2-\epsilon}\Gamma(1-2\epsilon)}
$
and the $A_i$ are defined through equations \eqref{eq1} - \eqref{a2}. We use \nobreak{$\mu_R=80$ GeV}; it enters our calculation through the usual
modification of the loop integration measure $\displaystyle \int {\rm
  d}^4 k \to \mu_{R}^{2\ep} \int {\rm d}^D k $ in dimensional
regularization.  The one-loop amplitudes are calculated in the
four-dimensional helicity scheme \cite{fdh}. In Table~\ref{tab2} we give the ratio S
\begin{equation}
S=\frac{4\pi}{\alpha_s}\frac{\text{Re}(A^{\text{one-loop}}A^{\text{tree}*})}{|A^{\text{tree}^2}|}
\end{equation}
of amplitudes summed over spin and color. Note that \textit{all} unrenormalized virtual amplitudes (including fermion loops) contribute to $A^{\text{one-loop}}$.

\begin{table}[h]
\begin{center}
\begin{tabular}{|l|c|c|c|}
\hline
Amplitude& $\; 1/\e^2 \;$ & $1/\e$ & $\e^0$ \\
\hline
$\;A_0(\udcs)$&& 
 & $\;  -7.488599-i\,    17.47659 
$
\\
$\;r_a(\udcs)$
&$\;-2.000000\;$
& $\; 5.580909+i\,    0.000000$ & $\;    11.010146+i\,  14.39329\;$\\
$\;r_a(\bar{u},d,s,\bar{c})$
&$\;-2.000000\;$
& $\; 6.634395+i\,    0.000000$ & $\;    -16.15227-i\,  7.909865\;$\\
$\;r_b(\uscd)$
&$\;-1.000000\;$
& $\; 3.551457-i\,    3.141593$ & $\;    -2.605243+i\,  9.184110\;$\\
$\;r_c(\uscd)$
&$\;-1.000000\;$
& $\; -2.092934-i\,    3.141593$ & $\;   -1.237780-i\, 9.355851\;$\\
$\;r_d(\uscd)$
&& $\; -0.666667+i\,    0.000000$ & $\; 1.431296 -i\,2.305401\;$\\
\hline 

\end{tabular}
\end{center}
\caption{
  Numerical results for the primitive tree-level amplitude $A_0(\udcs)$, 
  in units of $10^{-11}~{\rm GeV}^{-4}$ 
  and the ratios of primitive one-loop amplitudes $r_i$.
}
\label{tab1}
\end{table}

\begin{table}[h]
\begin{center}
\begin{tabular}{|l|c|c|c|}
\hline
Ratio& $\; 1/\e^2 \;$ & $1/\e$ & $\e^0$ \\
\hline
$\;S(\udcs)$
&$\;-5.333333\;$
& $\; 13.62554$ & $\;  23.35965 \;$\\
\hline

\end{tabular}
\end{center}
\caption{
  Numerical results for the ratio of spin and color summed amplitudes $S(\udcs)$.
}
\label{tab2}
\end{table}


\newpage
\newpage

\begin{thebibliography}{99}

\bibitem{dps}
A.~Kulesza and W.~J.~Stirling, 
Phys. Lett. B{\bf 475}, 168 (2000);\\
%
  E.~Maina,
  JHEP {\bf 0909}, 081 (2009);\\ 
J.~R.~Gaunt, C.~H.~Kom, A.~Kulesza and W.~J.~Stirling,
  hep-ph/1003.3953.


\bibitem{dreiner}H.~K.~Dreiner, S.~Grab, M.~Kr\"amer and M.~K.~Trenkel,
  Phys. Rev.  D{\bf 75}, 035003 (2007).
  
\bibitem{han}T.~Han, I.~Lewis and T.~McElmurry, 
JHEP {\bf 1001}, 123 (2010).

\bibitem{maalampi} See e.g. J.~Maalampi and N.~Romanenko, 
Phys. Lett.~B {\bf 532}, 202 (2002) and references therein. 

\bibitem{jager}B.~Jager, C.~Oleari and D.~Zeppenfeld, Phys. Rev.
  D{\bf 80}, 034022 (2009).

\bibitem{denner}
  A.~Denner and S.~Dittmaier,
  Nucl.\ Phys.\  B {\bf 734}, 62 (2006).

\bibitem{golem} T.~Binoth, J.~P. Guillet, G.~Heinrich, 
E.~Pilon and C.~Schubert, JHEP {\bf 0510}, 015 (2005).

\bibitem{britto}
  R.~Britto, F.~Cachazo and B.~Feng,
  Nucl.\ Phys.\  B {\bf 725}, 275 (2005).

\bibitem{Britto:2004tx}
  R.~Britto, F.~Cachazo and B.~Feng,
  Phys.\ Lett.\  B {\bf 611}, 167 (2005).


\bibitem{Forde:2007mi}
  D.~Forde,
  Phys.\ Rev.\  D {\bf 75}, 125019 (2007).

  
\bibitem{opp}G.~Ossola, C.~G.~Papadopoulos and R.~Pittau, Nucl. Phys.
  B~{\bf 763}, 147 (2007).

\bibitem{opp2}G.~Ossola, C.~G.~Papadopoulos and R.~Pittau, 
JHEP {\bf 0805}, 004 (2008).


\bibitem{egk}R.~K.~Ellis, W.~T.~Giele and Z.~Kunszt, JHEP {\bf 0803},
  003 (2008).
  
\bibitem{Giele:2008ve}W.~T.~Giele, Z.~Kunszt and K.~Melnikov, JHEP
  {\bf 0804}, 049 (2008).

\bibitem{cfb} C.~F.~Berger {\it et al.}, 
Phys. Rev. D {\bf 78}, 036003 (2008).
  
  
\bibitem{Bredenstein:2009aj}A.~Bredenstein, A.~Denner, S.~Dittmaier and
  S.~Pozzorini, Phys. Rev. Lett. {\bf 103}, 012002 (2009).

  
\bibitem{Bredenstein:2010rs}A.~Bredenstein, A.~Denner, S.~Dittmaier and
  S.~Pozzorini, JHEP {\bf 1003}, 021 (2010).
  
\bibitem{Bevilacqua:2009zn}G.~Bevilacqua, M.~Czakon, C.~G.~Papadopoulos,
  R.~Pittau and M.~Worek, JHEP {\bf 0909}, 109 (2009).
  
\bibitem{Berger:2009zg}C.~F.~Berger {\it et al.}, Phys. Rev. Lett.
  {\bf 102}, 222001 (2009).
  
\bibitem{Berger:2009ep}C.~F.~Berger {\it et al.}, Phys. Rev.
  D~{\bf 80}, 074036 (2009).
  
  

\bibitem{Ellis:2009zw}R.~K.~Ellis, K.~Melnikov and G.~Zanderighi, JHEP
  {\bf 0904}, 077 (2009).


\bibitem{KeithEllis:2009bu}R.~Keith Ellis, K.~Melnikov and G.~Zanderighi,
  Phys. Rev. D {\bf 80}, 094002 (2009).


\bibitem{Melnikov:2009wh}
K.~Melnikov and G.~Zanderighi, Phys. Rev. D  {\bf 81}, 074025 (2010).
  

\bibitem{bbbb}
T.~Binoth, N.~Greiner, A.~Guffanti, J.-Ph.~Guillet, T.~Reiter, J.~Reuter,
Phys.~Lett.~B {\bf 685}, 293, (2010).


\bibitem{tt2j} G. Bevilacqua, M. Czakon, C. G. Papadopoulos and M. Worek,
Phys.~Rev.~Lett. {\bf 104}, 162002 (2010).

\bibitem{Berger:2010vm}C.~F.~Berger {\it et al.}, arXiv:1004.1659
  [hep-ph].


\bibitem{Ellis:2008qc}R.~K.~Ellis, W.~T.~Giele, Z.~Kunszt, K.~Melnikov and
  G.~Zanderighi, JHEP {\bf 0901}, 012 (2009).
  
\bibitem{BG}F.A.~Berends and W.~T.~Giele, Nucl. Phys.~B
  {\bf 306}, 759 (1988).


\bibitem{cs} S.~Catani and M.~H.~Seymour, Nucl.~Phys. B{\bf 485}, 
291 (1997) [Erratum-ibid. B {\bf 510}, 503; 
(1998)].

  
\bibitem{nagy}
Z.~Nagy and Z.~Trocsanyi, 
 Phys.\ Rev.\  D {\bf 59}, 014020 (1999)
   [Erratum-ibid.\  D {\bf 62}, 099902 (2000)];
Z. Nagy, Phys. Rev. D~{\bf 68}, 094002 (2003).
  
\bibitem{mcfm}J.~M.~Campbell and R.K.~Ellis, Phys. Rev. D~{\bf 62},
  114012 (2000). The MCFM program is publicly available from
 http://mcfm.fnal.gov.

\bibitem{qcdloops} R.K.~Ellis and G.~Zanderighi, 
JHEP {\bf 0802}, 002 (2008).


\bibitem{Bern:1994fz}
  Z.~Bern, L.~J.~Dixon and D.~A.~Kosower,
  Nucl.\ Phys.\  B {\bf 437}, 259 (1995).

\bibitem{Bern:1997sc}
  Z.~Bern, L.~J.~Dixon and D.~A.~Kosower,
  Nucl.\ Phys.\  B {\bf 513}, 3 (1998).


\bibitem{rules} Z.~Bern, L.~Dixon and D.~Kosower, 
Ann. Rev. Nucl. Part. Sci. {\bf 46}, 109 (1996).

\bibitem{Ellis:2008ir} 
R.~K.~Ellis, W.~.T~Giele, Z.~Kunszt and K.~Melnikov, Nucl. Phys.~B {\bf 822}, 270 (2009).

\bibitem{Melnikov:2010iu}
K.~Melnikov and M.~Schulze, Nucl. Phys.~B {\bf 840}, 129 (2010).

    

\bibitem{Cacciari:2008gp}
M.~Cacciari, G.~P.~Salam and G.~Soyez,
JHEP {\bf 0804},  063 (2008). 

\bibitem{Martin:2009iq}
  A.~D.~Martin, W.~J.~Stirling, R.~S.~Thorne and G.~Watt,
  Eur.\ Phys.\ J.\  C {\bf 63},  189 (2009).


\bibitem{nlowork} Contribution by
S.~H\"oche, J.~Huston, D.~Maitre, J.~Winter and G.~Zanderighi
to T.~Binoth {\it et al.}
``The SM and NLO multileg working group: Summary report'',
arXiv:1003.1241[hep-ph].


\bibitem{eboli} A.~Alves, O.~Eboli, T.~Plehn and D.~Rainwater.
Phys. Rev. D~{\bf 69}, 075005 (2004).


\bibitem{fdh}  Z.~Bern and D.~A.~Kosower, Nucl. Phys.~B 
{\bf 379}, 451 (1992);  Z.~Bern, A.~De Freitas, L.~J.~Dixon and 
H.~L.~Wong, Phys. Rev.~D {\bf 66}, 085002 (2002).





\end{thebibliography}
\end{document}